\documentclass[twocolumn]{aastex62}
\usepackage{csvsimple}
\usepackage{amsmath}  
\usepackage{amssymb}  
\usepackage{multirow}
\usepackage{arydshln}
\usepackage{array}
\newcommand{\PreserveBackslash}[1]{\let\temp=\\#1\let\\=\temp}
\newcolumntype{C}[1]{>{\PreserveBackslash\centering}p{#1}}
\newcolumntype{R}[1]{>{\PreserveBackslash\raggedleft}p{#1}}
\newcolumntype{L}[1]{>{\PreserveBackslash\raggedright}p{#1}}

\newcommand{\kep}{{\it Kepler}}
\newcommand{\prot}{$P_\text{rot}$}
\newcommand{\sph}{$S_\text{\!ph}$}
\newcommand{\teff}{$T_\text{eff}$}
\newcommand{\logg}{$\log\, g$}
\usepackage{colortbl}
\definecolor{purple}{RGB}{102,0,204}
\newcommand\aastex{AAS\TeX}

\shorttitle{\aastex\ Rotation for Kepler M and K stars}
\shortauthors{A. R. G. Santos et al.}

\usepackage{hyperref}

\begin{document}

\title{\MakeUppercase{Surface rotation and photometric activity for \textit{Kepler} targets}
\MakeUppercase{I. M and K main-sequence stars}}

\author{A. R. G. Santos}
\email{asantos@spacescience.org}
\affil{Space Science Institute, 4750 Walnut Street, Suite 205, Boulder CO 80301, USA}

\author{R. A. Garc\'{i}a}
\affil{IRFU, CEA, Universit\'e Paris-Saclay, F-91191 Gif-sur-Yvette, France}
\affil{Universit\'e Paris Diderot, AIM, Sorbonne Paris Cit\'e, CEA, CNRS, F-91191 Gif-sur-Yvette, France}

\author{S. Mathur}
\affil{Instituto de Astrofisica de Canarias, Spain}
\affil{Universidad de La Laguna, Spain}

\author{L. Bugnet}
\affil{IRFU, CEA, Universit\'e Paris-Saclay, F-91191 Gif-sur-Yvette, France}
\affil{Universit\'e Paris Diderot, AIM, Sorbonne Paris Cit\'e, CEA, CNRS, F-91191 Gif-sur-Yvette, France}

\author{J. L. van Saders}
\affil{Institute for Astronomy, University of Hawai'i, Honolulu, HI 96822}

\author{T. S. Metcalfe}
\affil{Space Science Institute, 4750 Walnut Street, Suite 205, Boulder CO 80301, USA}
\affil{Max-Planck-Institut f\"ur Sonnensystemforschung, Justus-von-Liebig-Weg 3, 37077, G\"ottingen, Germany}

\author{G. V. A. Simonian}
\affil{Department of Astronomy, The Ohio State University, 140 West 18th Avenue, Columbus, OH 43210}

\author{M. H. Pinsonneault}
\affil{Department of Astronomy, The Ohio State University, 140 West 18th Avenue, Columbus, OH 43210}

\begin{abstract}
Brightness variations due to dark spots on the stellar surface encode information about stellar surface rotation and magnetic activity. 
In this work, we analyze the \kep\ long-cadence data of 26,521 main-sequence stars of spectral types M and K in order to measure their surface rotation and photometric activity level.
Rotation-period estimates are obtained by the combination of a wavelet analysis and autocorrelation function of the light curves. Reliable rotation estimates are determined by comparing the results from the different rotation diagnostics and four data sets. We also measure the photometric activity proxy \sph\ using the amplitude of the flux variations on an appropriate timescale.
We report rotation periods and photometric activity proxies for about 60 per cent of the sample, including 4,431 targets for which \citet{McQuillan2013a,McQuillan2014} did not report a rotation period. For the common targets with rotation estimates in this study and in \citet{McQuillan2013a,McQuillan2014}, our rotation periods agree within 99 per cent. In this work, we also identify potential polluters, such as misclassified red giants and classical pulsator candidates.
Within the parameter range we study, there is a mild tendency for hotter stars to have shorter rotation periods. The photometric activity proxy spans a wider range of values with increasing effective temperature. The rotation period and photometric activity proxy are also related, with \sph\ being larger for fast rotators. Similar to  \citet{McQuillan2013a,McQuillan2014}, we find a bimodal distribution of rotation periods.
\end{abstract}

%

\section{Introduction}

Stellar rotation is a key ingredient for the generation of magnetic fields and magnetic cycles in the Sun and other solar-type stars \citep[e.g.][]{Brun2017}. Stars are observed to spin down as they evolve and lose angular momentum \citep[e.g.][]{Wilson1963,Wilson1964}. Therefore, rotation can also be used as a diagnostic for stellar age, in what we call gyrochronology \citep[ e.g.][]{Skumanich1972,Barnes2003,Barnes2007,Mamajek2008,Garcia2014,Davies2015,Metcalfe2019}. To calibrate the empirical gyrochronology relations, rotation-period estimates for large samples stars are needed as well as precise ages estimates. 

Thanks to the NASA mission \kep, almost 200,000 stars were observed almost continuously for up to four years. These long duration photometric observations obtained with high precision allow us to measure rotation periods through the modulation of the stellar brightness caused by the passage of spots on the stellar disk. This has been done on a large number of stars observed by {\it Kepler} using different techniques, such as periodogram analysis \citep[e.g.][]{Nielsen2013, Reinhold2013a}, autocorrelation function \citep[e.g.][]{McQuillan2013a, McQuillan2014}, and time-frequency analysis with wavelets \citep[e.g.][]{Garcia2014}. Based on simulated data, the comparison of different pipelines developed to retrieve surface rotation periods using photometric data showed that a combination of different techniques such as done in \citet{Garcia2014} and \citet{Ceillier2016, Ceillier2017} provides the most complete and reliable set of rotation-period estimates \citep[see details in][]{Aigrain2015}.

Ages can be constrained from gyrochronology relations. However, these relations are calibrated, requiring targets with known ages from independent methods. This is the reason why stars belonging to clusters have been used in the past \citep[e.g.][]{Meibom2011a,Meibom2011b,Meibom2015}. Asteroseismology has proven to be a powerful tool to provide precise stellar ages \citep[e.g.][]{Mathur2012,Metcalfe2014,SilvaAguirre2015,Creevey2017,Serenelli2017} and we can now test and improve those relations with a large number of field stars. This led to the results of \citet{vanSaders2016} who found that solar-like stars older than the Sun rotate faster than predicted by the classical gyrochronology relations \citep[see also][]{Angus2015}. The authors suggested that this could be the result of the weakening of the magnetic braking when the Rossby number of the star (ratio between the rotation period and the convective turnover time) reaches a given value. Thus, gyrochronology may not be a uniformly suitable technique for all main-sequence stars, and the apparent weakened braking has implications for the dynamo theory. \citet{vansaders2018} suggested that signatures of this weakened braking might be visible in large samples of field stars with measured rotation periods. However, these relations are still valid for young main-sequence stars. Note that gyrochronology is also not suitable for pre-main-sequence and early spectral type stars \citep[e.g.][]{Kraft1967,Gallet2013,Epstein2014,Amard2016}.

One of the largest analyses of the surface rotation of main-sequence stars observed by \kep\ was performed by \citet{McQuillan2014}. They looked for rotation periods in a sample of 133,030 targets using Quarters 3 to 14 and obtained reliable values for 34,030 stars. They estimated ages by comparing their field population to empirical gyrochrones and found that the most slowly rotating stars were consistent with a gyrochronological age of 4.5 Gyr. 

In this work we perform a similar analysis focusing on 26,521 M and K dwarfs observed by \kep. We use the longest time-series available: up to \kep\ Quarter 17, while \citet{McQuillan2014} used only 11 \kep\ Quarters. We calibrate the light curves using our own independent software, which high-pass filters the data using filters of 20, 55, and 80 days, preventing us from measuring a harmonic of the real rotation period. In addition, we analyze the PDC-MAP \citep[Presearch Data Conditioning - Maximum A Posteriori; e.g][]{Jenkins2010,Smith2012,Stumpe2012} light curves to be sure that the rotational modulation detected does not result from photometric pollution by nearby stars. We are particularly careful to remove possible polluters such as red giants, classical pulsators, and eclisping binary systems, which can result in the detection of a spurious periodicity. The sample selection and data calibration are described in Sect.~\ref{sec:datasample}. We then apply our rotation pipeline (Sect.~\ref{sec:method}) that consists of the combination of the auto-correlation function, the wavelet analysis, and the composite spectrum (that is a combination of the two former methods) to derive the most reliable periods. Our analysis allows us to retrieve rotation periods for more than 4,000 additional targets in comparison with the analysis in \citet{McQuillan2014}. In particular, we are able to retrieve rotation periods for both fainter and cooler stars. We then measure the photometric magnetic activity proxy $S_{\rm ph}$ (Sect.~\ref{sec:sph}). Finally we interpret the results in Sect.~\ref{sec:res} in terms of rotation, activity, mass, and temperature relations and conclude 

in Sect.~\ref{sec:conclusion}.

\section{Data preparation and sample selection}\label{sec:datasample}

\subsection{Data preparation}\label{sec:data}

The light curves are obtained from \kep\ pixel-data files using large custom apertures that produce stable light curves. For each pixel in the pixel-data file, a reference flux value is computed as the 99.9\textsuperscript{th} percentile of the flux. Starting from the center of the point-spread function of the target-pixel mask, new pixels are added in one direction of the mask if two conditions are fulfilled: 1) the reference flux of the pixel is higher than a threshold of $100 \text{e}^-\!/\text{s}$; 2) the flux is smaller than the value of the previous pixel within a small tolerance. In most cases, this second condition allows us to remove the pixels corresponding to a second star present in the aperture.

The resulting light curve is processed through the implementation of the \kep\ Asteroseismic Data Analysis and Calibration Software \citep[KADACS;][]{Garcia2011}. KADACS corrects for outliers, jumps, and drifts, and it properly concatenates the independent \kep\ Quarters on a star-by-star basis. It also fills the gaps shorter than 20 days in long-cadence data following in-painting techniques based on a multi-scale cosine transform \citep{Garcia2014a,Pires2015}. The resulting light curves are high-pass filtered at 20, 55 days (quarter by quarter) and 80 days (using the entire light curve) yielding three different light curves for each target. For light curves longer than one month, KADACS corrects for discontinuities at the edges of the \kep\ Quarters. Hereafter, we will refer to KADACS data products, which are optimized for seismic studies, as KEPSEISMIC\footnote{KEPSEISMIC time-series are available at MAST via \dataset[https://doi.org/10.17909/t9-cfke-ps60]{https://doi.org/10.17909/t9-mrpw-gc07}.}.

To correctly prepare the KOI (\kep\ Objects of Interest) light curves for seismic analysis, we used the published ephemeris of each star from the MAST (Mikulski Archive for Space Telescopes) to remove the transits and interpolate the resultant gaps using the same in-painting techniques mentioned above.

In the analysis below we compare the results for the different high-pass filters to determine the final stellar rotation period. However, we note that the longer period filters are also less effective at removing long periodicity instrumental trends from the light curves, including the \kep\ yearly modulation. 

In our analysis, to ensure that the correct rotation period is retrieved, we also use PDC-MAP light curves for Data Release 25. The comparison between KEPSEISMIC and PDC-MAP light curves also helps to identify light curves with photometric pollution by nearby stars. We typically construct the KEPSEISMIC light curves using larger apertures than those of the PDC-MAP time-series, which leads to an increase in the number of polluted light curves. Nevertheless, a significant number of light curves show evidence of pollution or multiple signals in both KEPSEISMIC and PDC-MAP data sets (see Sect.~\ref{sec:sample}, and Tables~4 and 5). To properly understand the source of pollution and/or multiple signals (e.g. possible binary or just nearby star in the field of view) further analyses are needed and currently beyond the scope of this study. Note that, in addition to the difference in the aperture sizes, PDC-MAP light curves are calibrated quarter-by-quarter while the KEPSEISMIC light curves are calibrated using all the quarters at once. Furthermore, PDC-MAP light curves are often high-pass filtered at a period of 20 days which can lead to biased results (see Appendix \ref{app2}).

\subsection{Sample selection}\label{sec:sample}

\begin{figure}[h]
\includegraphics[width=\hsize]{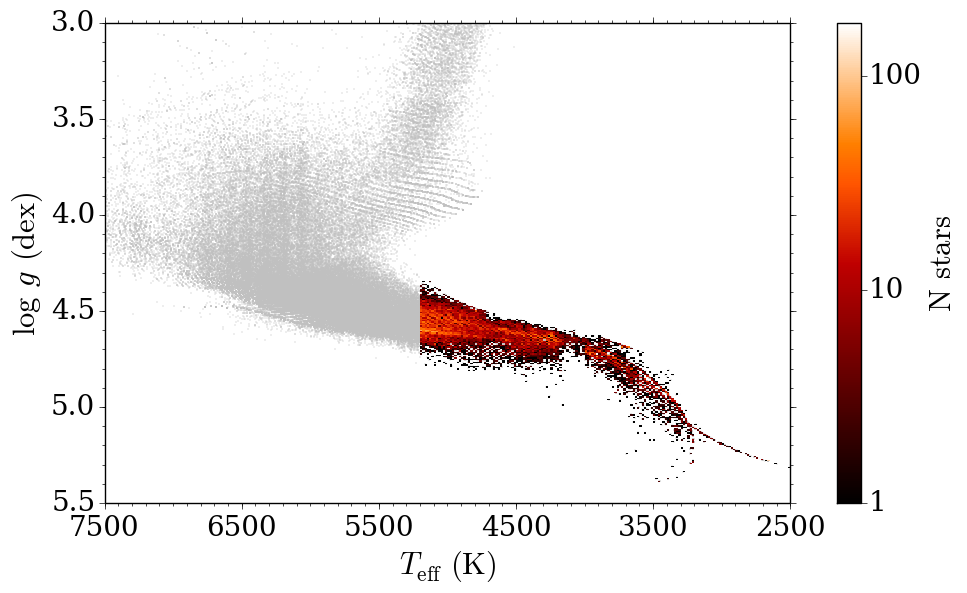}
\caption{Surface gravity-effective temperature diagram of the 26,521 M and K dwarfs according to KSPC DR25 \citep{Mathur2017}, color coded by number of stars in each bin. The size of \teff\ and \logg\ bins is $\sim16$ K and $\sim 7\times10^{-3}$ dex, respectively. For context, other stars in KSPC DR25 are plotted in gray. Effective temperature (\teff) and surface gravity (\logg) values are adopted from KSPC DR25.}\label{fig:hrsample}
\end{figure}

We analyze long-cadence ($\Delta t=29.42\,\rm min$) data of M and K main-sequence stars observed during the main mission of the \kep\ satellite \citep{Borucki2010}. The targets were selected according to the \kep\ Stellar Properties Catalog for Data Release 25 \citep[KSPC DR25;][]{Mathur2017}, where M~dwarfs have effective temperatures \teff\ smaller than 3700 K and K~dwarfs have temperatures within 3700 and 5200 K.
The initial sample is composed by 26,521 targets (24,171 K stars and 2350 M stars) shown in Fig.~\ref{fig:hrsample}.

We expect a number of different polluters in the sample of main-sequence M and K stars. These polluters will display stellar variability due to pulsations, eclipses or other astrophysical variability not related to spot-modulation. Therefore, we search and identify such polluters.

We start by removing the known eclipsing binaries \citep[total of 272 stars in the Villanova \kep\ Eclipsing Binary Catalog;][]{Kirk2016,Adbul-Masih2016} and known RR Lyrae (3 stars in this sample; Szab\'o et al. in prep). These are listed in Table~4. For rotational analysis of eclipsing binaries see \citet{Lurie2017}.

We also remove possible misclassified red giants (listed in Table~4; Garc\'ia et al. in prep). A significant fraction of those were identified by \citet{Berger2018} using astrometric data from \textit{Gaia} \citep[\textit{Gaia} Data Release~2;][]{Gaia_DR2}. The remainder of the misclassified targets were identified using the hallmark signature of red-giant stars in light curves: the presence of red-giant oscillations. We use both neural network and machine learning techniques that automatically identify power spectra consistent with red-giant stars \citep[see][]{Hon2018,Bugnet2018} and/or by visual examination for red-giant pulsations. In total, 1,221 misclassified red giants were removed from the subsequent rotation analysis. 30 of those targets are also identified as eclipsing binaries (flagged accordingly in Table~4), which may suggest that one of the components of the binary is a red giant.

Another group of potential polluters in the sample corresponds to light curves exhibiting evidence of photometric pollution possibly from nearby stars in the field of view. We consider light curves to be photometrically polluted when the signal is only present in some \kep\ Quarters, namely every four Quarters\footnote{Every $\sim90$ days, i.e. every \kep\ Quarter, the spacecraft was rotated over $90^\circ$, meaning that the targets are observed by the same modules/channels every four \kep\ Quarters.}. We also identify targets as photometrically polluted when their light curves contain a signal or multiple signals only in the KEPSEISMIC time-series. As mentioned previously, the apertures used for the KEPSEISMIC data sets are typically larger than those of the PDC-MAP data sets, and thus more likely to be affected by the contribution of background stars in the field of view. In total, we have flagged the light curves of 255 targets as photometrically polluted (Table~4). 

Targets with multiple signals in both the PDC-MAP and KEPSEISMIC light curves are likely to be associated with different unresolved sources. Although determining whether these targets are true binary systems or merely polluted by background stars is beyond the scope of this work, we perform the rotation analysis of these light curves (Table~5). In total, we have identified 270 targets with multiple signals in both KEPSEISMIC and PDC-MAP data sets.

Another concern is pollution by possible classical pulsators (CP) that were not previously identified. We start by flagging the targets that exhibit multiple high-amplitude peaks at relatively high frequencies (higher than $3.5 \mu\text{Hz}$) in the power density spectrum, which are typical of classical pulsators. Then we visually check those targets and also all the other targets for which the rotation estimate (from Sect.~\ref{sec:a2z}) is shorter than 10 days. We only flag stars as CP candidates when there are more than three associated peaks in the power spectra. 

\begin{figure*}[htp]
\includegraphics[width=\hsize]{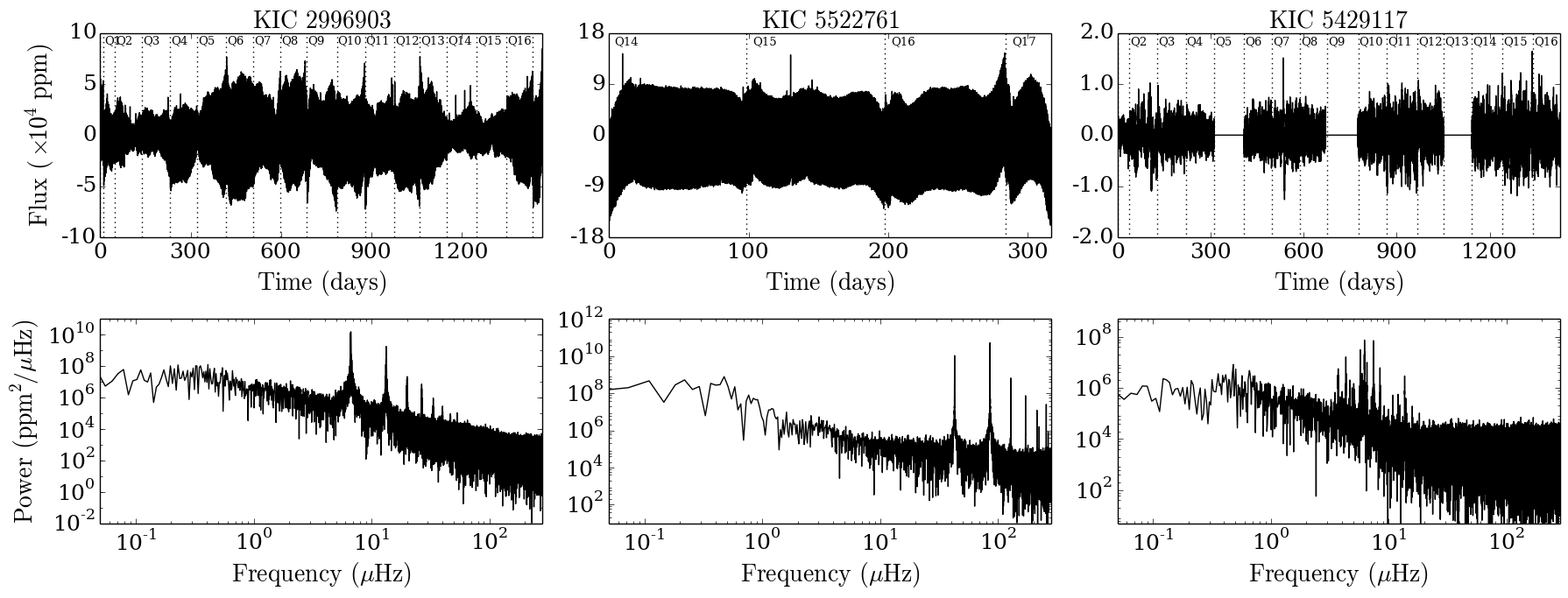}
\caption{Light curve and power density spectrum for an example of the three classical pulsator or close-in binary candidates. {\it Left-hand}: KIC~2996903, Type~1 CP/CB candidate, which exhibit high-amplitude flux variations and high-amplitude peaks with a large number of harmonics in the power density spectrum. {\it Middle}: KIC~5522761, Type~2 CP/CB candidate, which exhibit high-amplitude flux variations and a large number of high-amplitude peaks, with the highest peak being the second harmonic of the signal period. {\it Right-hand}: KIC~5429117, Type~3 CP candidate, which is possibly a $\gamma$ Doradus. Note that targets marked as Type~3 CP candidates may be $\gamma$ Doradus or $\delta$ Scuti depending on the nature of the modes (and characteristic frequencies).}\label{fig:cp}
\end{figure*}

We also distinguish between three types of CP candidates. Type~1 candidates (left panels of Fig.~\ref{fig:cp}) show a behaviour somewhat similar to RR Lyrae and Cepheids \citep[see e.g. Szab\'o et al. in prep;][]{Kolenberg2010,Moskalik2015}: high-amplitude and stable flux variations, beating patterns, and a large number of harmonics. Interestingly, a significant fraction of these targets were identified as \textit{Gaia} binary candidates in \citet{Berger2018} and \citet{Simonian2019}. In particular, \citet{Simonian2019} focused on tidally synchronized binary systems. Of the 74 Type~1 candidates we identify common to their analysis, 51 are found to be possible synchronized binaries. Therefore, it is possible that these targets are not classical pulsators but close-in binaries (CB). If that is the case, the signal may still be related to rotation, but may be distinct from the  rotational behavior of single stars. For the remainder of this paper, we refer to these targets (350) as Type~1 CP/CB candidates. Type~1 CP/CB candidates are listed and flagged in Table~3. Targets marked as Type~2 CP/CB candidates (9 stars; middle panel of Fig.~\ref{fig:cp}) exhibit a large number of harmonics in the power spectrum, similarly to Type~1 CP/CB candidates. However, these targets differ from Type~1, in particular, the highest peak in the periodogram is the second harmonic associated with the signal instead of the first harmonic (period of the signal). This signature may also be consistent with contact binary systems \citep[see e.g.][]{Lee2016,Colman2017}. Therefore, similarly to Type~1, these targets are flagged as CP/CB candidates. The power spectrum of Type~3 CP candidates (9 stars; right-hand panel of Fig.~\ref{fig:cp}) resembles those of $\gamma$ Doradus or $\delta$ Scuti, depending on characteristic frequencies and nature of the modes \citep[see e.g.][]{Bradley2015,vanReeth2015,BarceloForteza2017}. A proper analysis of these targets is however beyond the scope of this work. Type~2 and 3 candidates are listed in Table~4. 

We do not provide rotation periods for confirmed RR Lyrae, misclassified red giants, eclipsing binaries, light curves with photometric pollution, and Type~2 and 3 CP/CB candidates. This leaves us with 24,782 stars for the rotational analysis. Table~\ref{tab0} summarizes the number of polluters and targets used in the subsequent analysis.

\begin{table}[h]
    \centering
    \begin{tabular}{rl}
    \hline\hline
    M dwarfs & 2,156\\
    K dwarfs & 22,006 \\
    Type~1 CP/CB candidates & 350\\
    Multiple signals & 270\\\hline
    Eclipsing Binaries (EB)& 242\\
    Red giants (RG) & 1,191\\
    EB \& RG & 30\\
    RR Lyrae & 3\\
    Photometric pollution & 255\\
    Type 2 and 3 CP/CB candidates & 18\\\hline\hline
    \end{tabular}
    \caption{Summary of the targets classified as M and K~dwarfs in KSPC DR25 \citep{Mathur2017}. The top part of the table corresponds to the targets for which we perform the rotational analysis, while the polluters summarized in the bottom part are not used for rotational analysis.}
    \label{tab0}
\end{table}

Finally, possible additional non-single non-main-sequence M and K stars are flagged in Tables~3-5 but not removed from the analysis. We add subgiant and binary flags from \citet[][\textit{Gaia} DR2]{Berger2018}, synchronized binary flag from \citet{Simonian2019}, and FliPer$_\text{Class}$ flag \citep[see][]{Bugnet2019} which indicates solar-type stars, classical pulsators, and binary/photometric pollution. We do not remove these targets from the analysis, but we alert for the possible pollution.

\section{Surface rotation detection}\label{sec:method}

In Sect.~\ref{sec:a2z}, we present the methodology implemented to estimate the surface rotation period. Sections~\ref{sec:auto} and \ref{sec:vis} summarize the results from the automatic selection and visual examination, respectively.

\subsection{Methodology to retrieve rotation periods}\label{sec:a2z}

To extract the rotation-period estimates, we implement the methodology described in \citet{Ceillier2016,Ceillier2017}. It combines a time-frequency analysis and the autocorrelation function (ACF). This methodology was found by \citet{Aigrain2015} to have the best performance in terms of completeness and reliability compared to the periodogram analysis alone, ACF alone, or a combination between the two and spot modeling. 

Despite the fact that our KEPSEISMIC light curves have been corrected for instrumental effects \citep[see Sect.~\ref{sec:data};][]{Garcia2011}, calibrated light curves may still exhibit instrumental modulations. We therefore remove \kep\ Quarters with anomalously high variance compared with their neighbours from the rotation analysis \citep[see][]{Garcia2014}.

\begin{figure}[h!]
\includegraphics[width=\hsize]{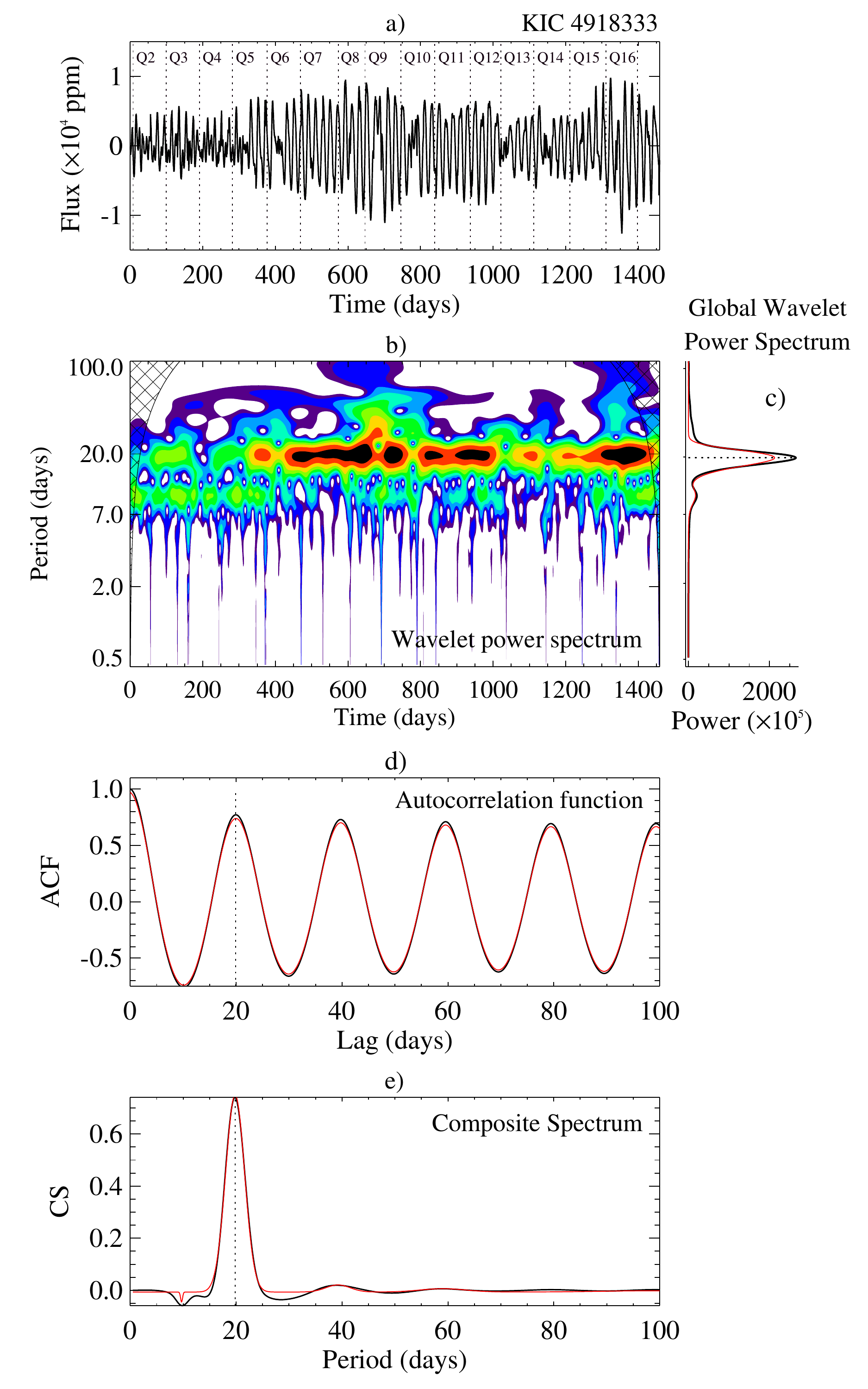}
\caption{Rotation analysis for KIC~4918333. a) KEPSEISMIC light curve obtained with the 55-day filter. b) Wavelet power spectrum (WPS) where red and black correspond to high power and blue to low power. The cone of influence is shown by the black crossed area. c) Global wavelet power spectrum (GWPS; black) and corresponding best fit with multiple Gaussian functions (red). d) Autocorrelation function (ACF; black) of the light curve and smoothing ACF (red). e) Composite spectrum (black) and respective fit with multiple Gaussian profiles (red). For the GWPS, ACF, and CS, the black dotted lines mark the respective rotation-period estimates.}\label{fig:a2ztool}
\end{figure}

First, we estimate periods from a time-period analysis using the wavelet decomposition \citep{Torrence1998} adapted by \citet{Mathur2010b} using the correction by \citet{Liu2007}. The wavelet analysis assesses the correlation between the mother wavelet and the rebinned data (to decrease the computing time) by sliding the wavelet in time for a given period of the wavelet. The range of periods is probed through an iterative process. For the mother wavelet, we use the Morlet wavelet, which is the convolution between a sinusoidal and a Gaussian function. This analysis provides the wavelet power spectrum (WPS). An example is given in panel b) of Fig.~\ref{fig:a2ztool}, where red and black colors indicate high power, while blue indicates low power. The visual inspection of the WPS also helps us to determined whether the signal is present along the time-series or an artifact resulting from instrumental noise at a particular time. The black hashed area indicates the cone of influence that marks the limit on observing at least four rotations in the light curve. Rotation signals found inside the cone have a lower confidence level. Finally, we obtain the global wavelet power spectrum (GWPS) by computing the sum of the WPS along time for each period of the wavelet (panel c) of Fig.~\ref{fig:a2ztool}. We then fit the GWPS, through a least-squares minimization, with multiple Gaussian functions. The rotation estimate from the wavelet analysis corresponds to the central period of the highest fitted period peak, while the uncertainty corresponds to the half width at half maximum (HWHM) of the corresponding Gaussian profile. Computed in this way, the inferred uncertainty also accounts for possible differential rotation.

Our second method for measuring periods consists of the autocorrelation function of light curves \citep[ACF; following the procedure in ][]{McQuillan2013}, which was combined with wavelet analysis for the first time in \citet{Garcia2014}. The ACF is smoothed using a Gaussian function whose width is a tenth of the most significant period selected from the Lomb-Scargle periodogram \citep{Lomb1976,Scargle1982} of the ACF. We identify the significant peaks and take the highest peak as the rotation-period estimate from the ACF. We also examine the ACF for evidence of double-peaked features resulting from active regions in anti-phase. Panel d) of Fig.~\ref{fig:a2ztool} shows the ACF for a given target in the sample.

Finally, the third method of estimating rotation period utilizes the composite spectrum (CS), which combines the GWPS and the ACF as described by \citet{Ceillier2016,Ceillier2017}. The composite spectrum corresponds to the product of the normalized GWPS and the normalized ACF resampled in the same period of the GWPS. Periods present in both methods, GWPS and ACF, are enhanced by the CS, allowing for a better identification of the intrinsic rotation periods of the star. We also fit the CS with multiple Gaussian functions; the central period and HWHM of the profile corresponding to the highest peak are taken as its period estimate and uncertainty. Panel e) of Fig.~\ref{fig:a2ztool} shows an example of the CS.

For the rotation-period estimate provided in Tables~3 and 5, we prioritize the value returned by the wavelet analysis. When the wavelet power spectrum does not allow us to successfully recover the rotation period, we provide the value recovered from the composite spectrum. If both wavelet power spectrum and composite spectrum fail to infer the rotation period, the rotation period provided is that found by the autocorrelation function without uncertainty. Note that the primary goal of the autocorrelation function and composite spectrum is to validate the rotation period and better identify the reliable results.

\subsubsection{Automatic selection}\label{sec:auto}

Having the rotation-period estimates from the GWPS, ACF, and CS for the three sets of KEPSEISMIC light curves, we start by selecting the targets with the most reliable rotation estimates.

For the automatic selection, the appropriate filter is chosen according to the rotation period. We note that it is still possible to recover periods longer than the cut-off period of the filter. The transfer function is unity below the cutoff period. Above that, it varies sinusoidally and slowly approaches zero at twice of the cut-off period. Therefore, the amplitude of rotation periods slightly longer than the cut-off period would be only slightly reduced, while rotation periods close to 1.5 times the cut-off period would have roughly half of the original amplitude. It is therefore still possible to extract a high signal-to-noise ratio, reliable peak of a period 1.5 times the cut-off period using our rotation pipeline. However, the 80-day filtered light curves are the least stable often exhibiting instrumental modulations and, thus, we only use them in the automatic selection for rotation periods longer than 60 days. For rotation periods shorter than 23 days, priority is given to the period estimate obtained from the 20-day filter. For rotation periods between 23 and 60 days, the primary filter is the 55-day filter, while for longer periods priority is given to the 80-day filter. 

The targets with reliable rotation-period estimates are automatically selected if:
\begin{enumerate}
  \item for a given filter, the rotation-period estimates from GWPS, ACF, and CS agree within $2\sigma$ where $\sigma$ is chosen to be the period uncertainty from GWPS;
  \item the rotation-period estimates agree within $20\%$ between different filters
  \begin{enumerate}
      \item for \prot$<60$ days, the rotation estimates agree between the 20-day and 55-day filters;
      \item or for \prot$\ge60$ days, the rotation estimates agree between the 55-day and 80-day filters;
  \end{enumerate}
  \item for the appropriate filter, the peak height in the ACF and CS are larger than a given threshold. We adopt the thresholds imposed by \citet{Ceillier2017}:
  \begin{enumerate}
    \item $G_\text{ACF}\ge0.2$, where $G_\text{ACF}$ is the height of the ACF peak which corresponds to \prot;
    \item $H_\text{ACF}\ge0.3$, where $H_\text{ACF}$ is the mean difference between the height of the ACF peak and the values of the two local minima on either side of the peak;
    \item and $H_\text{CS}\ge0.15$, where $H_\text{CS}$ is calculated in the same manner as $H_\text{ACF}$ but for the CS.
  \end{enumerate}
\end{enumerate}

Following the steps above, 9,586 targets were automatically selected (this number includes Type~1 CP/CB candidates), which corresponds to $\sim60\%$ of the total number of targets for which we provide rotation-period estimates (Table~3). Targets whose retrieved period is consistent with the reported orbital periods for confirmed and candidate planet hosts (data from the Exoplanet Archive) are reported as targets with no spot modulation (Table~4).

\subsubsection{Visual Check}\label{sec:vis}

For stars that were not automatically selected we proceed to visually check their KEPSEISMIC (three filters) and PDC-MAP light curves, the respective power density spectra, and the rotation diagnostics. We also visually check the light curves of the targets for which the rotation results for the PDC-MAP light curves are not consistent with those for the KEPSEISMIC light curves. Often, half of the rotation period is recovered from the PDC-MAP time-series (see Appendix \ref{app2}). This is probably due to the fact that PDC-MAP applies a 20-day filter, but not systematically in all quarters or to all stars. The comparison with PDC-MAP also helps to identify KEPSEISMIC light curves polluted by nearby stars, as the latter use larger apertures (see Sect.~\ref{sec:data}). Targets showing evidence for photometric pollution are listed in Table~4.

Although multiple signals present in both the KEPSEISMIC and PDC-MAP light curves may still be the result of photometric pollution by background stars, we determine and report the periods of the observed multiple signals (Table~5). Note that these multiple signals are most likely not related to differential rotation as the detected periods are well separated. For most of these targets, the periods of the different signals have to be determined through visual inspection and manually, for example, by limiting the range of period to be searched. For some of the targets, one of the multiple signals is consistent with one of the CP/CB candidates described above. Thus, we also provide the respective flag in Table~5. For signals consistent with Type~2 and 3 we do not provide a period. Finally, we note that some of the signatures can be the result of eclipses or transits. Although some of the targets with multiple signals are KOIs reported as false negatives, none of these targets is a confirmed eclipsing binary.

The rotation-period estimate in Tables~3 and 5 is provided as described in Sect. \ref{sec:a2z}, prioritizing the results from the wavelet analysis.

From the visual inspection, the rotation periods for 6,324 additional targets were determined. In total, we provide rotation-period estimates for 15,910 targets (Tables~3-5; including Type~1 CP/CB candidates and light curves with multiple signals).

Although a significant number of targets exhibit evidence for rotational modulation (3,562), we are not able to confidently recover rotation periods. Generally, their light curves exhibit instrumental effects, which hamper the detection of the true rotation period. We mark these in Table~4 as targets with possible spot modulation. 

From this analysis, we find that 5,310 targets (also listed in Table~4) show no evidence for spot modulation. This could be due to the combination of small amplitude spot modulation and noise, or due to the spot visibility, which depends on the stellar inclination angle and spot latitudinal distribution.

\begin{table}[h]
    \centering
    \begin{tabular}{R{1.5cm}cc}
    \hline\hline
    \multicolumn{3}{c}{\bf With $\pmb{P}_\text{\!rot}$ estimate}\\\hline
    & Auto. selected & Visually selected  \\
    M dwarfs & 918 & 612 \\
    K dwarfs & 8,380 & 5,380 \\\hline
    \multicolumn{3}{c}{Type~1 CP/CB candidates\hspace{0.2cm}350} \\\hline
    \multicolumn{3}{c}{Multiple signals\hspace{0.2cm}270} \\\hline\\
    
    \multicolumn{3}{c}{\bf Without $\pmb{P}_\text{\!rot}$ estimate}\\\hline
    & No rotation & Possible rotation\\
    M dwarfs & 494 & 132\\
    K dwarfs & 4,816 & 3,430\\\hline\hline
    \end{tabular}
    \caption{Summary of the results from the rotational analysis of $24,782$ targets. The top part of the table corresponds to the targets for which we provide \prot\ estimate (automatically and visually selected). The bottom part of the table corresponds to the targets for which we do not provide \prot\ estimate. Part of those targets do not exhibit spot modulation, while others show possible spot modulation but we are unable to confidently provide a \prot\ value.}
    \label{tab01}
\end{table}\pagebreak

\section{Photometric magnetic activity proxy}\label{sec:sph}

Using CoRot \citep[Convection, Rotation, and planetary Transits;][]{Baglin2006} data for the solar-type star HD~49933, \citet{Garcia2010} showed that the light curve variability due to the presence of magnetic features on the stellar surface --- including starspots --- provides a proxy of stellar magnetic activity. However, brightness variations may include contributions from different phenomena, such as active regions, granulation, oscillations, stellar companions, or instrumental effects. Different phenomena affect the light curve at different timescales. Therefore, to properly estimate a photometric magnetic activity proxy, the stellar rotation period must be taken into account. \citet{Mathur2014} determined that the activity proxy \sph\ computed as the standard deviation of sub-series of length $5\times P_\text{rot}$ provides a reasonable measure of activity and is primarily related to magnetism and minimizes the contributions from other sources of variability. Furthermore, using VIRGO \citep[Variability of Solar Irradiance and Gravity Oscillations][]{Frohlich1995} and GOLF \citep[Global Oscillations at Low Frequency][]{Gabriel1995} data, the photometric activity proxy \sph\ was shown to recover the variation associated with the solar activity cycle at both 11-year and quasi-biennial timescales \citep{Salabert2017}.
For seismic solar-analog stars observed by \kep\ and by the ground-based, high-resolution \textsc{Hermes} spectrograph \citep{Raskin2011}, \citet{Salabert2016a} demonstrated that \sph\ measurements are consistent with the chromospheric activity index measured from the Ca K-line emission \citep{Wilson1978}.


Thanks to the \kep\ space mission, the photometric activity can be easily estimated through \sph\ for a large number of stars with known rotation periods (which we estimate here). This is a clear advantage in relation to chromospheric activity indexes, which require a large amount of ground-based telescope time and are only possible to measure for bright targets. However, the photometric variability depends on the visibility of active regions. For example, assuming a similar latitudinal distribution of active regions in the Sun for other solar-type stars (note that it may not be true for late-type M dwarfs), \sph\ will correspond to a lower limit of the true photometric activity level for stars with small inclination angle, i.e the angle between the rotation axis and the line of sight, which is unknown for most targets.

In this work, we compute the photometric activity index \sph\ for the M and K~dwarfs with period estimates obtained in Sect.~\ref{sec:method}. In Tables~3 and 5, the \sph\ value and respective uncertainty are provided as the mean value and standard deviation of the \sph\ computed over sub-series of length $5\times P_\text{rot}$. The \sph\ index is corrected for the photon noise following the approach by \citet{Jenkins2010}. However, for $1\%$ of the targets with \prot\ estimate, the correction from \citet{Jenkins2010} leads to negative \sph\ values. For such targets, the correction to the photon noise is instead computed from the flat component in the power density spectra. In Table~5 (light curves with multiple signals), the multiple \sph\ values for a given target may change significantly as they are computed at different timescales depending on the respective period.

\section{Results}\label{sec:res}

Following the methodology described in Sect.~\ref{sec:method}, surface rotation periods were successfully measured for 15,290 stars ($\sim62\%$ of the targets for which we perform the rotational analysis; 1,530 M and 13,760 K~dwarfs), and for additional 350 Type 1 CP/CB candidates and 270 targets whose light curves show multiple signals. The photometric activity proxy \sph\ was also measured for the same targets.

Tables~3 and 5 summarize the properties and results for the stars with rotation-period estimates, including Type 1 CP/CB candidates and those lightcurves with multiple signals in both KEPSEISMIC and PDC-MAP data sets. Table~4 lists the remainder of the target sample.



\begin{figure}[h]
\includegraphics[width=\hsize]{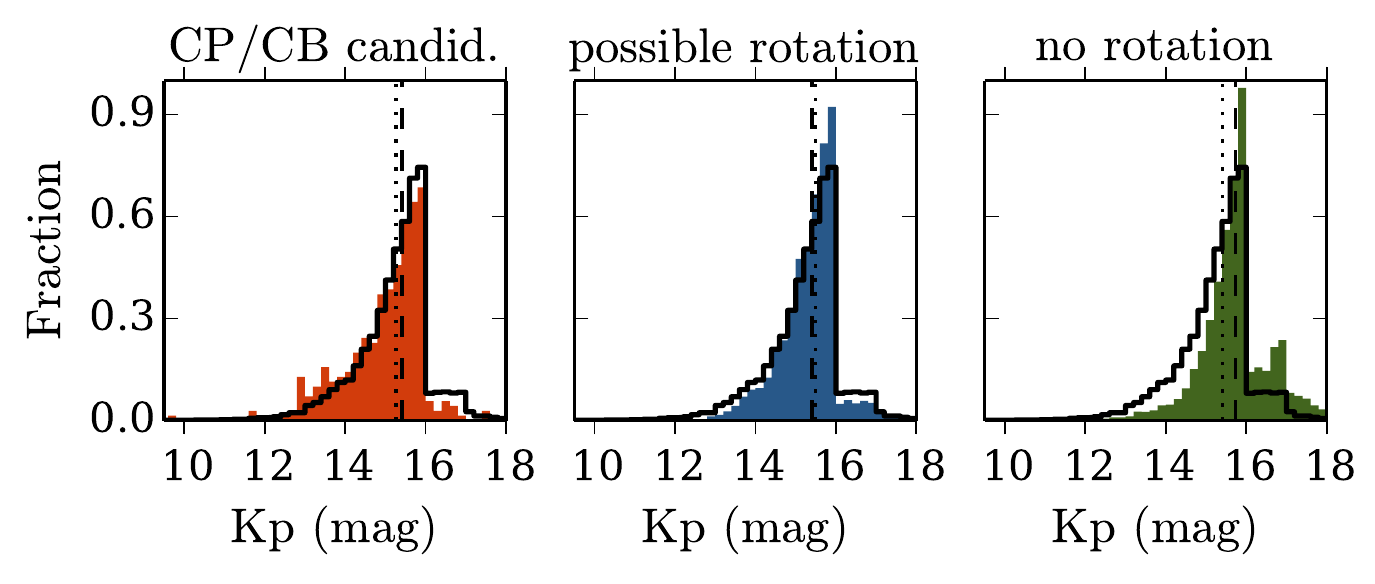}\vspace{-0.1cm}
\caption{Comparison between the magnitude distribution for stars with \prot\ estimate (excluding CP/CB candidates; black solid line) and that for: CP/CB candidates (left; red), stars with possible spot modulation (middle; blue), and stars without spot modulation (right; green). Dashed lines indicate the median magnitude for stars with \prot\ estimate, while the dotted lines correspond to the median value of the distributions shown in color.}\label{fig:kp}
\end{figure}

Figure~\ref{fig:kp} compares the distribution of \kep\ magnitudes (Kp) for targets with rotation-period estimate with those for CP/CB candidates (Type 1, 2, and 3), targets with possible spot modulation, and targets without evidence for spot modulation. The magnitude of stars with possible rotation modulation and of CP/CB candidates is consistent with that of stars with successful rotation measurement. For CP/CB candidates, there is, however, a slight excess of brighter targets. The distribution of targets that do not exhibit spot modulation extends to fainter magnitudes than that of targets with rotation estimates. Faint targets often show high levels of noise which hampers detection of rotational signatures.

\begin{figure}[h]
\includegraphics[width=\hsize]{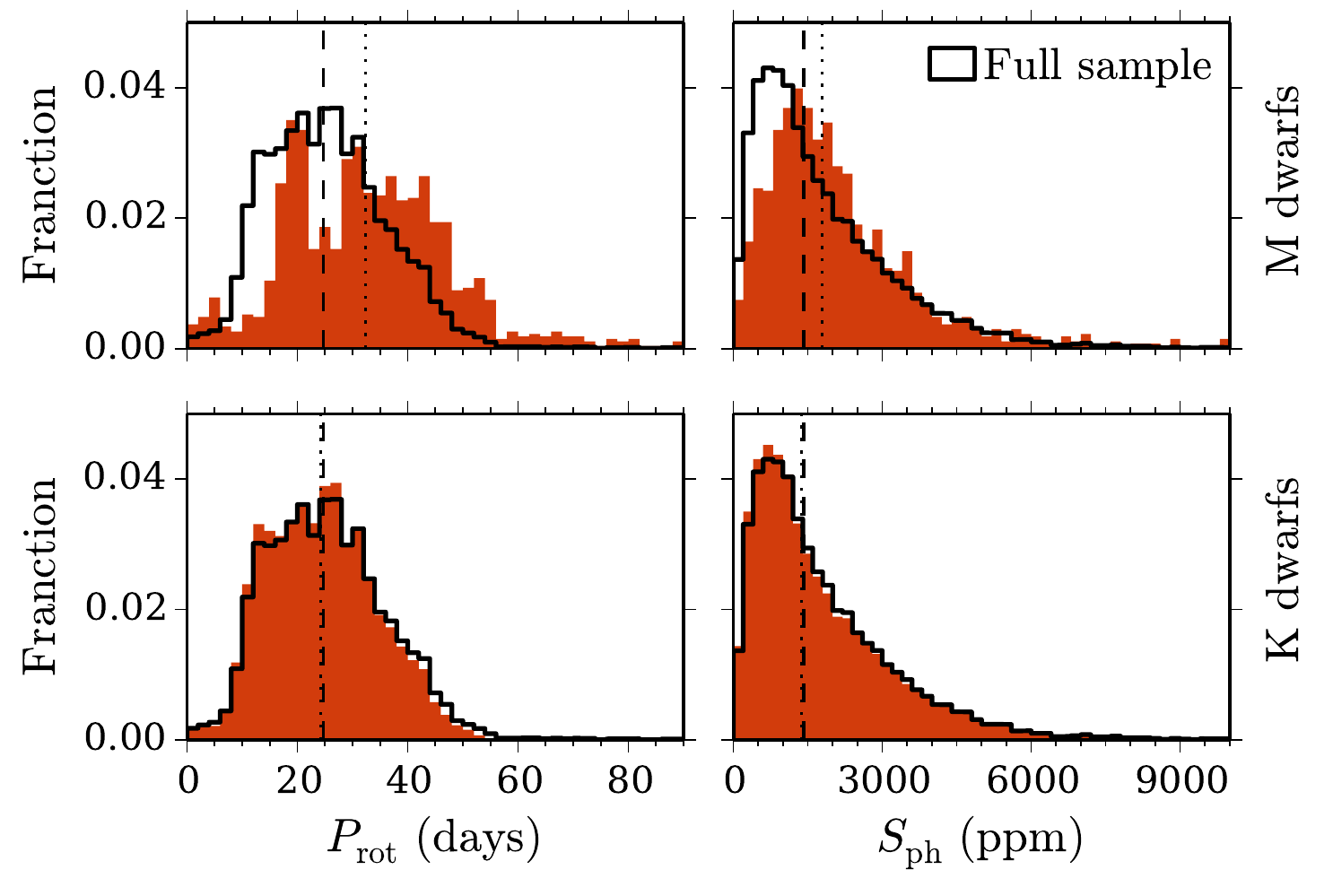}\vspace{-0.1cm}
\caption{Distribution of rotation periods (left) and \sph\ values (right) for M (top) and K~dwarfs (bottom) is shown in red. The respective median values are marked by the dotted lines. The distributions and corresponding median value for the full subsample of M anf K~dwarfs with \prot\ estimate are shown by the black solid and dashed lines, respectively.}\label{fig:hist}
\end{figure}

\begin{figure*}[htp]
\includegraphics[width=0.503\hsize]{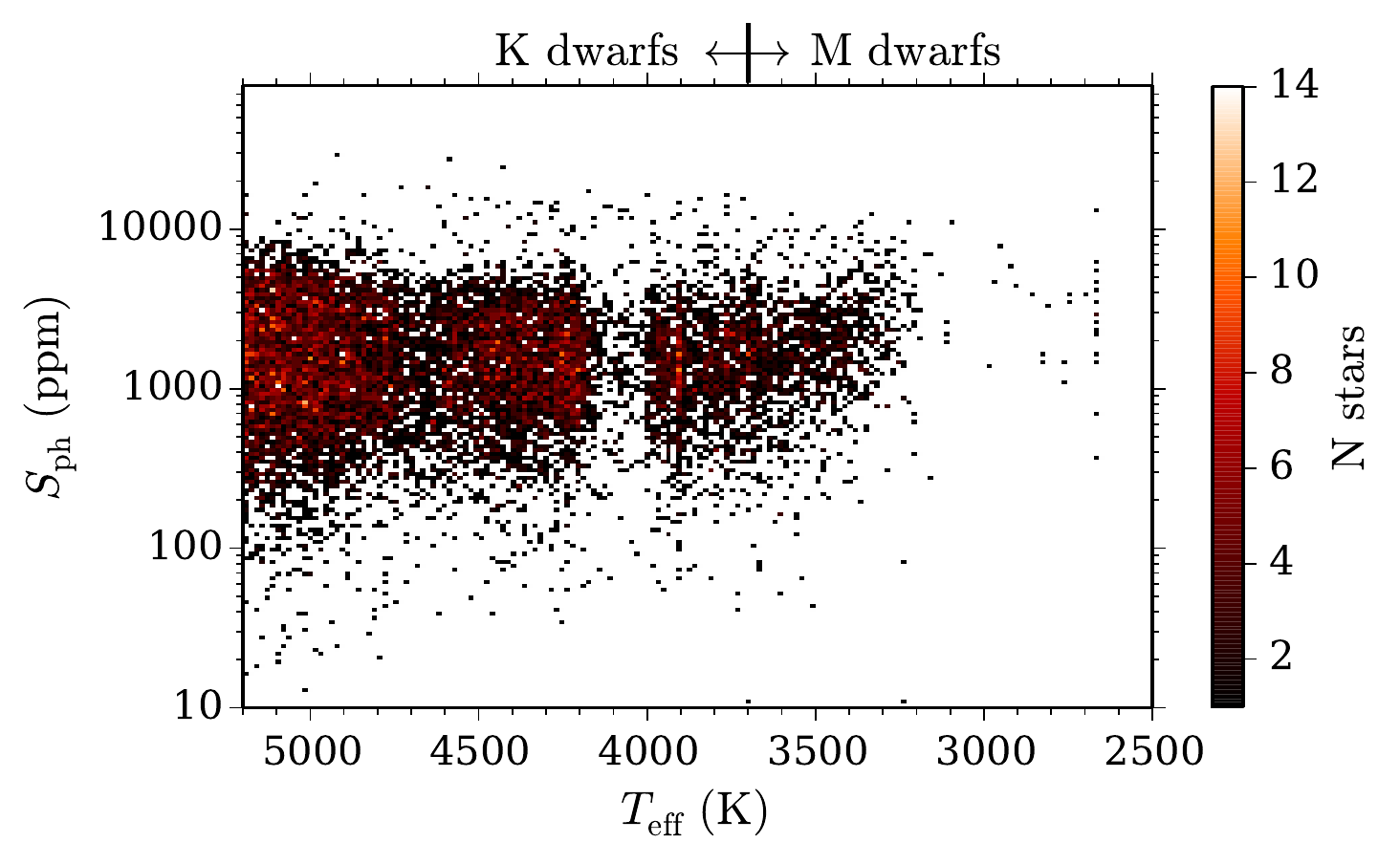}
\includegraphics[width=0.5\hsize]{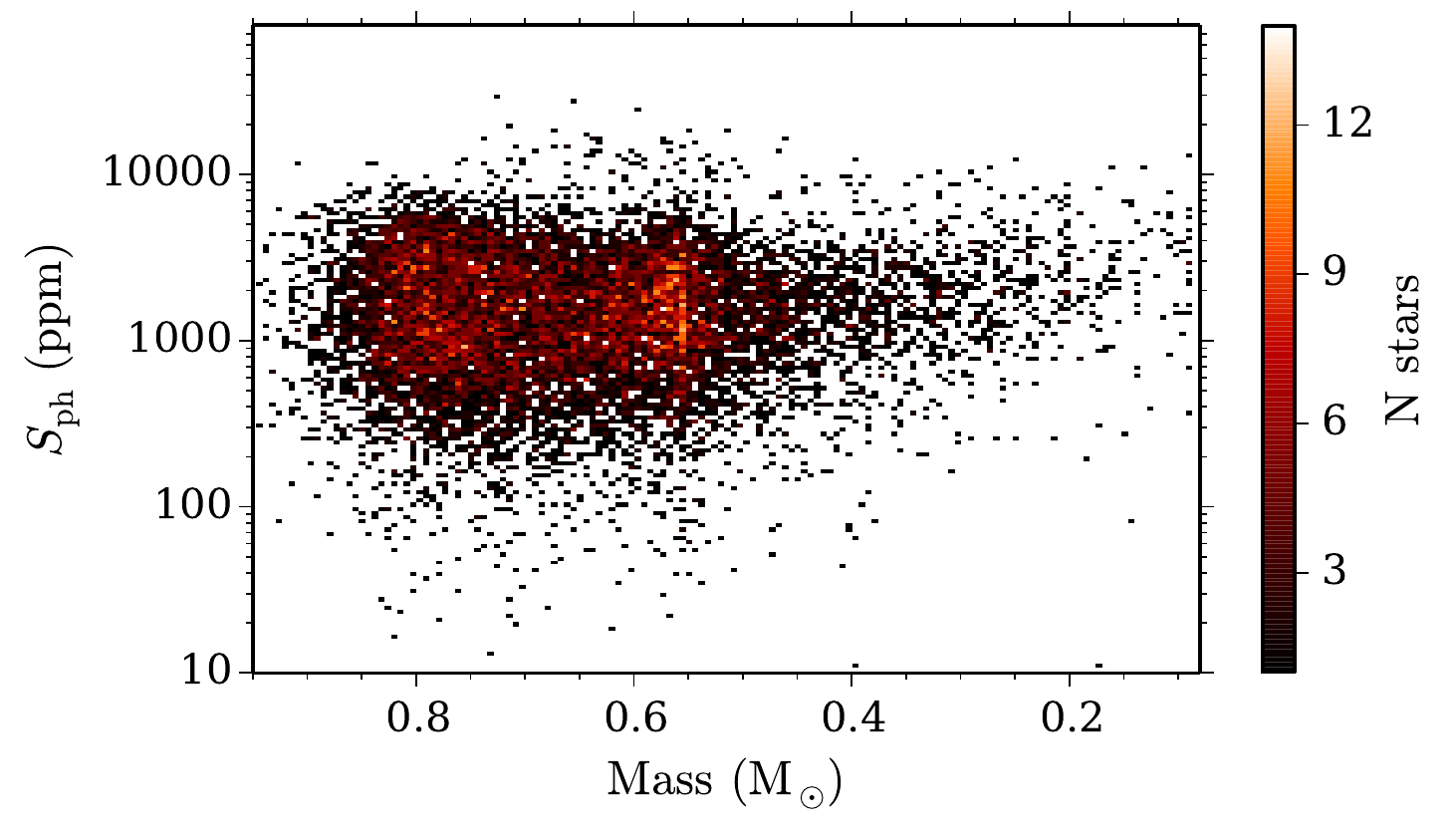}
\caption{Photometric activity index \sph\ as a function of effective temperature (left) and mass (right) color coded by number of stars in a given parameter range. Stellar effective temperature and mass are taken from KSPC DR25.} \label{fig:sphTeffMass}
\end{figure*}

Type~1 CP/CB candidates and targets whose light curves show multiple signals are neglected in Figs.~\ref{fig:hist}-\ref{fig:protsph} as well as all the possible non-single non-main-sequence stars flagged by \citet{Berger2018}, \citet{Simonian2019}, and FliPer$_\text{Class}$ \citep{Bugnet2019}. In Appendix \ref{app}, we present the same figures where all targets in Tables~3 and 5 are considered.

Figure~\ref{fig:hist} summarizes the results for the targets with period estimate. M~dwarfs have on average longer rotation periods and larger \sph\ values than K~dwarfs, which is consistent with the results in \citet{McQuillan2014}.

In the following sections, we take a more detailed look at the dependency of the surface rotation and photometric activity on the stellar effective temperature and mass.

\subsection{Rotation - Mass/Temperature relation}

\begin{figure*}[htp]
\includegraphics[width=0.503\hsize]{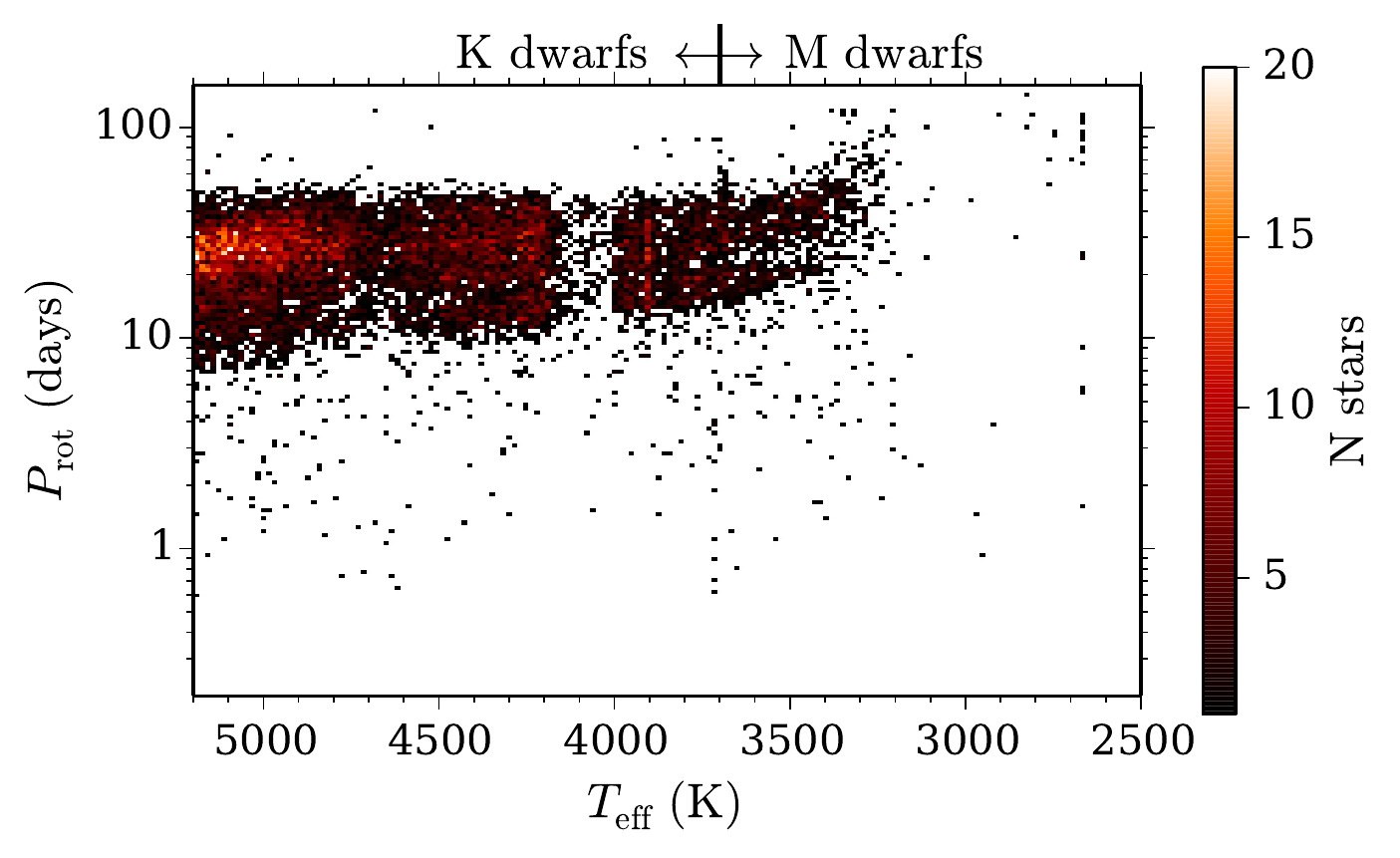}
\includegraphics[width=0.5\hsize]{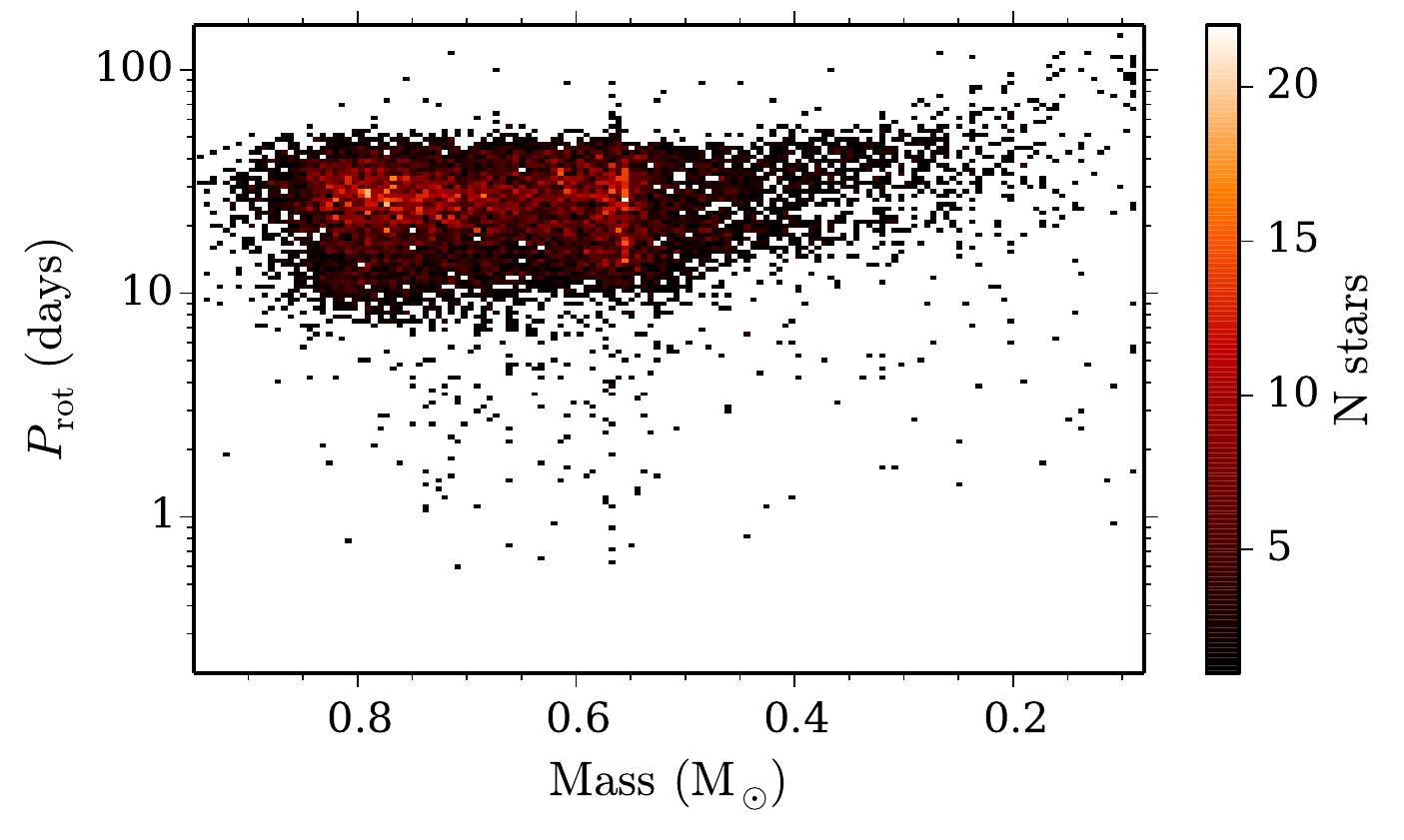}
\caption{Rotation period as a function of effective temperature (left) and mass (right) color coded by number of stars in a given parameter range. Brighter colors indicate higher density regions than darker colors. Stellar effective temperature and mass are taken from KSPC DR25. } \label{fig:protTeffMass}
\end{figure*}

The left-hand panel of Fig.~\ref{fig:protTeffMass} shows the rotation period as a function of stellar effective temperature (from KSPC DR25). As the effective temperature increases the average rotation period is found to decrease, meaning that hotter stars are generally faster rotators than cooler stars. Our results exhibit two sequences in the \prot-\teff\ relation which are consistent with the bimodal \prot\ distribution previously reported by \citet{McQuillan2013,McQuillan2014}. The vertical features and gaps are the result of artifacts in the \kep\ Stellar Properties Catalog temperature scale.

\begin{figure}\includegraphics[width=\hsize]{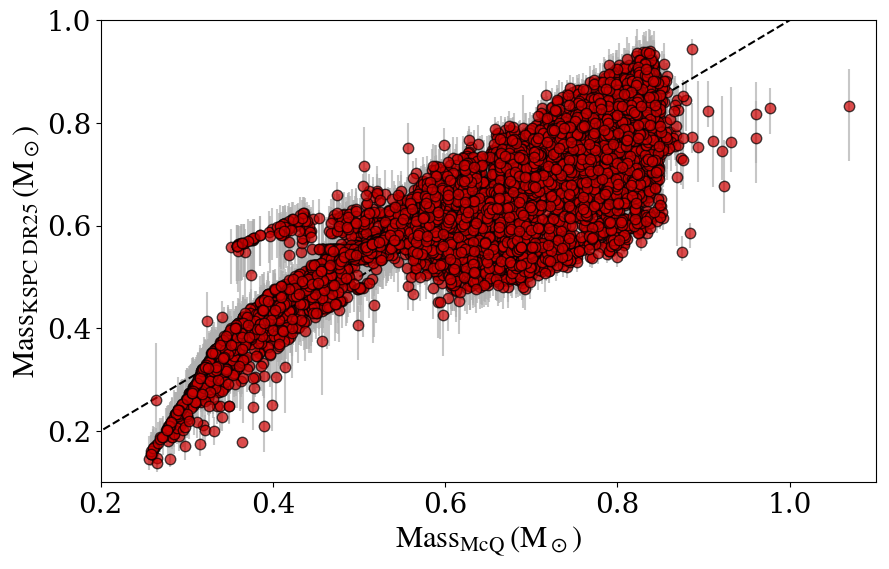}
\caption{Comparison between the stellar masses from \citet[][$\text{Mass}_\text{McQ}$]{McQuillan2014} and \citet[][$\text{Mass}_\text{KSPC DR25}$]{Mathur2017}.} \label{fig:mass}
\end{figure}

The right-hand side of Fig.~\ref{fig:protTeffMass} shows the rotation period as a function of stellar mass \citep[from KSPC DR25;][]{Mathur2017}. Rotation period decreases slightly with increasing mass. In this case, the bimodal rotation-period distribution is not as obvious as that in the \prot-\teff\ relation and, in particular, not as clear as in the \prot-mass relation found by \citet{McQuillan2014}. We note that the stellar masses used in this work are different from those in \citet{McQuillan2014}, as different stellar evolution codes with different physics and observables were used. As shown in Sect.~\ref{sec:mcq}, for the common targets, the period estimates obtained in this study and in \citet{McQuillan2014} are in very good agreement. Thus, the stellar masses are the source for the discrepancy. See Fig.~\ref{fig:mass} for the comparison between the masses from \citet{McQuillan2014} and those from \citet{Mathur2017}. 

\subsection{Photometric activity - Mass/Temperature relation}

The left-hand panel of Fig.~\ref{fig:sphTeffMass} shows the photometric activity proxy \sph\ as a function of the effective temperature. For the parameter space considered in this work, the photometric activity proxy takes on a wider range of values with increasing effective temperature. The upper envelope of \sph\ values increases with increasing temperature, while the lower envelope decreases. A similar behaviour is found for the \sph\ as a function of mass (right-hand panel in Fig.~\ref{fig:sphTeffMass}). Our results are consistent with those of \citet{McQuillan2014}.

The transition between fully convective stars and stars with a radiative core is expected to take place at $0.35\text{M}_\odot$ \citep[e.g.][]{Chabrier1997}. If the tachocline (transition between a differentially rotating convective envelope and a uniformly rotating radiative core) played an important role in the dynamo mechanism for M~stars, one might expect to observe a transition in the rotation period and photometric activity proxy distributions. However, due to the small number of targets with lower masses, there is no sufficient evidence to support or reject that hypothesis.

\subsection{Photometric activity - rotation relation}

\begin{figure}[h]
\includegraphics[width=\hsize]{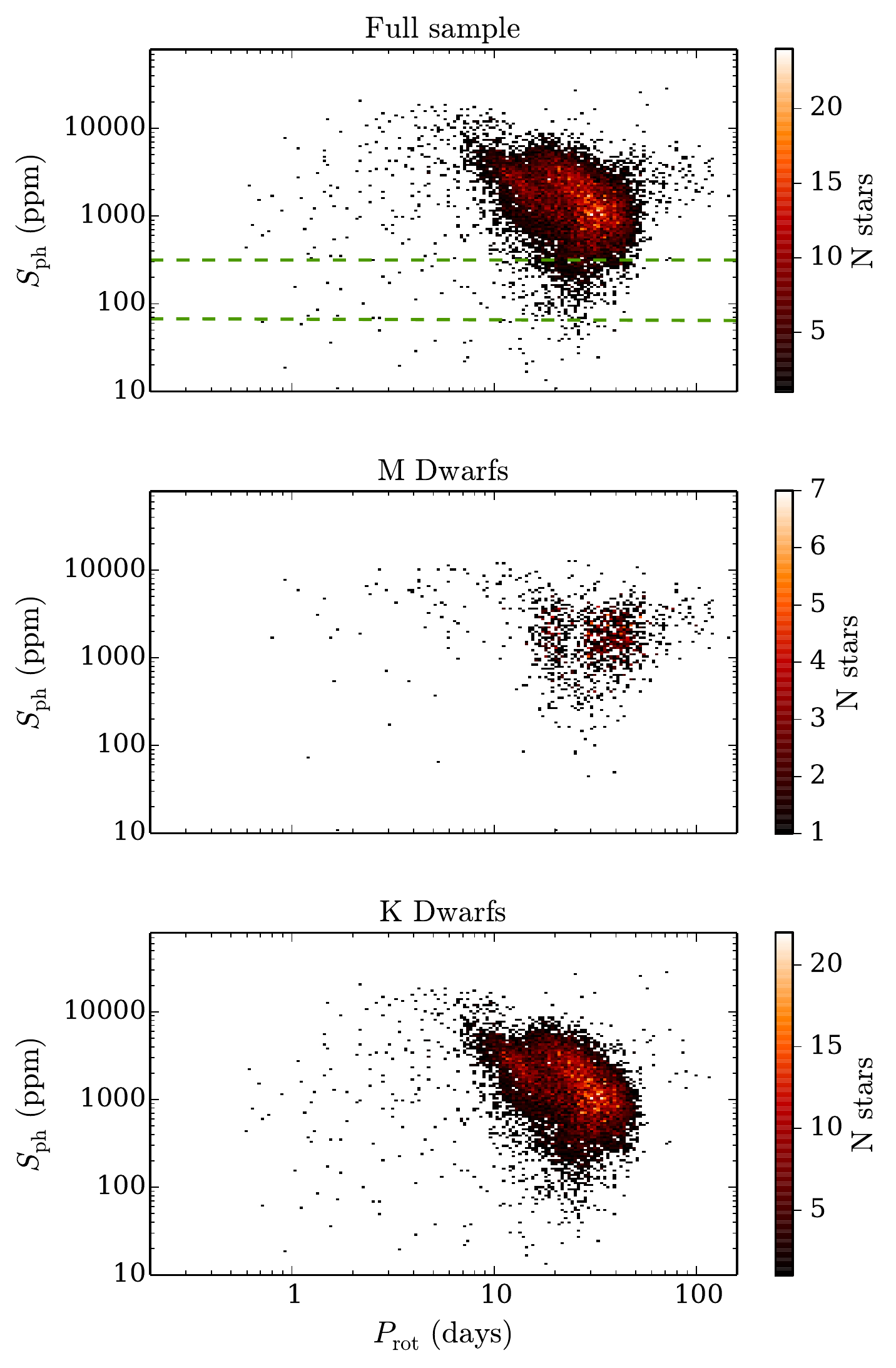}
\caption{Photometric activity proxy as a function of the rotation period color coded by number of stars in a given parameter range for: all M and K~dwarfs (top), M~dwarfs (middle), and K~dwarfs (bottom). For comparison, the \sph\ values at solar activity maximum (314.5 ppm) and minimum (67.4 ppm) are marked by the dashed green lines. Solar values from \citet{Mathur2014}.}\label{fig:protsph}
\end{figure}

Faster rotators are expected to be more active than slower rotators at fixed effective temperature \citep[e.g.][]{Vaughan1981,Baliunas1983,Noyes1984b}. Therefore, one should expect the photometric activity proxy \sph\ to be related to the rotation period. Figure~\ref{fig:protsph} shows the \sph\ as a function of the rotation period. For M~dwarfs there is no clear relationship. Nevertheless, for K~dwarfs we find a negative correlation: photometric activity increases with increasing rotation rate. The bimodality in the rotation-period distribution is also obvious in the \sph-\prot\ relation, which exhibits two distinct sequences for faster and slower rotators. Although we have made an effort to identify classical pulsator candidates, we note that there is still the possibility for additional polluters, namely Type~1 CP/CB candidates. We advise caution in particular when dealing with fast rotators with very large \sph\ values. Despite their similarity with targets flagged as Type~1 CP/CB candidates, these targets show three or less harmonics in the power spectra and thus do not obey the criteria imposed in Sect. \ref{sec:sample} to discriminate the CP/CB candidates.

\pagebreak

\subsection{Comparison with McQuillan et al. (2013, 2014)}\label{sec:mcq}

In this section, we compare our results with those from \citet{McQuillan2013a,McQuillan2014}. The periods from those works were estimated from the autocorrelation function of the PDC-MAP light curves for \kep\ Quarters 3-14 (3 years of data).

\begin{figure}[h]
\includegraphics[width=\hsize]{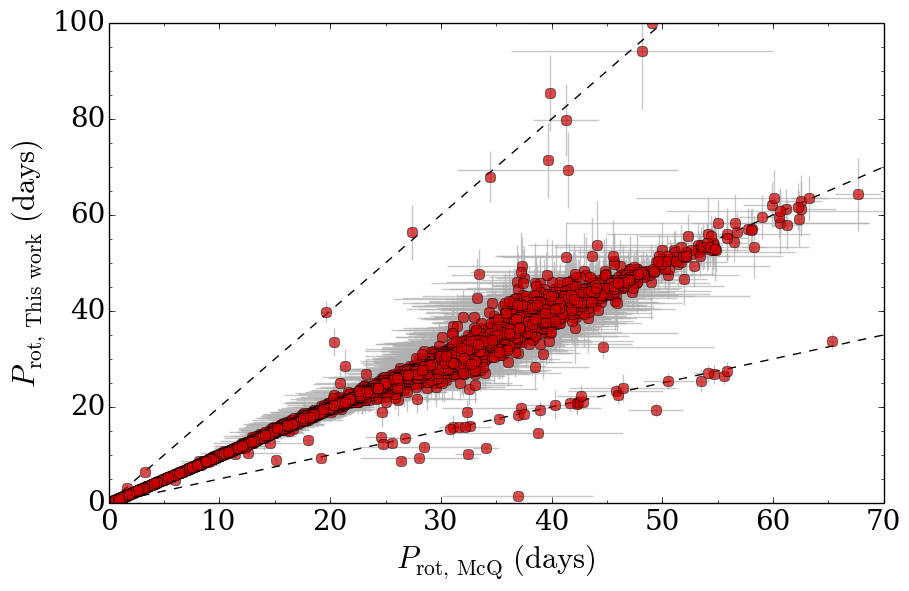}
\caption{Comparison between the rotation estimates from this work ($P_\text{rot,\,This\, work}$) and those from \citet[][$P_\text{rot,\,McQ}$]{McQuillan2013a,McQuillan2014}. The dashed lines indicate the two-to-one, one-to-one, and one-to-two lines.}\label{fig:McQ}
\end{figure}

Only 11,209 of the targets for which we provide a rotation-period estimate in Table~3 (15,640 targets in total) are common to the detection by \citet{McQuillan2013a,McQuillan2014}.
Figure~\ref{fig:McQ} shows the comparison between the rotation-period estimates, which agree within $2\sigma$ for $\sim99.4\%$ of the common targets ($\sim99.1\%$ within $1\sigma$). The most common cases outside of $2\sigma$ ($\sim0.3\%$ of the common targets) correspond to targets for which \citet{McQuillan2013a,McQuillan2014} measured double the rotation period found in our analysis. For a smaller number of stars ($\sim0.1\%$), \citet{McQuillan2013a,McQuillan2014} recovered half of the rotation period.

We provide rotation-period estimates for 4,431 targets (3,831 K stars; 618 M stars) that were not reported by \citet{McQuillan2013a,McQuillan2014}. From those only 558 targets were not identified as M and K main-sequence stars by \citet{Brown2011} and \citet{Dressing2013}, which provided the \kep\ properties adopted by \citet{McQuillan2013a,McQuillan2014}. Therefore, most of the additional estimates are rotation periods that \citet{McQuillan2013a,McQuillan2014} could not detect with their data and methodology.

\citet{McQuillan2014} reported rotation-period estimates for 465 targets in Table~4, including misclassified red giants (184), eclipsing binaries (5), RR Lyrae (3), Type~2 CP/CB candidates (2), and Type~3 CP candidates~(3). \citet{McQuillan2014} reported rotation periods for 26 targets that we have identified as not showing rotational modulation and 180 targets with possible rotational modulation. The time-series of these targets show significant instrumental effects which would lead to incorrect period estimates. The time-series of 62 targets for which \citet{McQuillan2014} reported \prot\ exhibit photometric pollution in both PDC-MAP and/or KEPSEISMIC data sets.

We also note that 286 targets with \prot\ estimate in \citet{McQuillan2014} are flagged as Type~1 CP/CB candidates in Table~3, while 179 show  multiple signals (Table~5) which can be related to multiple systems or photometric pollution by background stars. For the latter, an automatic rotation estimate will be biased towards the signal with largest amplitude.

\begin{figure}[h]
    \centering
    \includegraphics[width=\hsize]{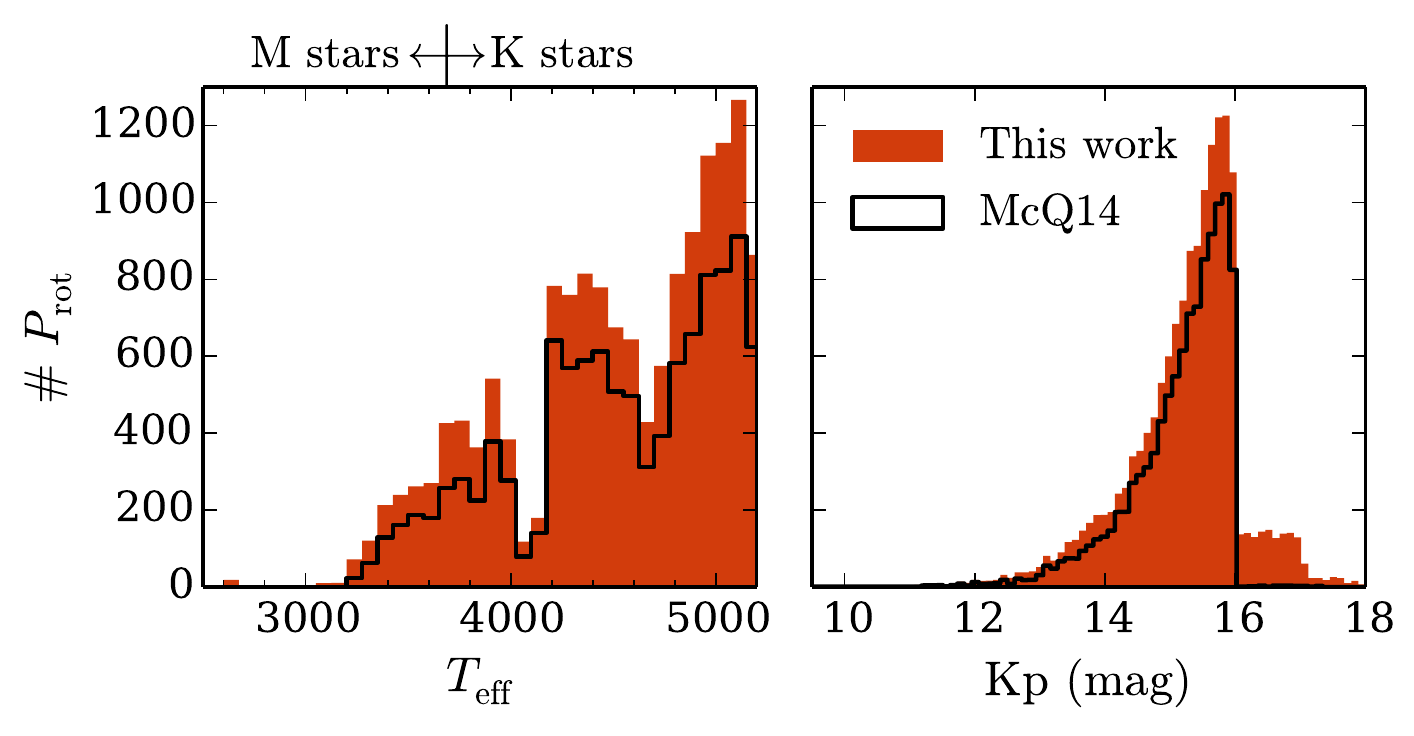}
    \caption{Distribution of the number of \prot\ estimates from this work (red) and from the analysis in \citet[][black]{McQuillan2014} as a function of effective temperature (left) and \kep\ magnitude (right).}\label{fig:histMcQ}
\end{figure}

The rotation analysis we perform combines the wavelet analysis with the autocorrelation function of the light curves \citep[e.g.][]{Garcia2014,Ceillier2016,Ceillier2017}. This methodology performs better than the autocorrelation function alone \citep[e.g.][]{McQuillan2013,McQuillan2013a,McQuillan2014} in terms of completeness and reliability \citep{Aigrain2015}. Furthermore, we used both the longest time-series available and our own calibrated light curves (KEPSEISMIC), which may have contributed to the significant improvement in the fraction of rotational signals we detect (see also Appendix~\ref{app2}). Figure \ref{fig:histMcQ} shows the comparison between the number of estimates in this work and in \citet{McQuillan2014}. As mentioned previously, we provide rotation periods for a larger number of stars --- in particular, the fraction of stars cooler than 4200 K with measured rotation periods is larger than that in \citet{McQuillan2014}.  Moreover, our methodology is also able to retrieve rotation periods for fainter stars than the analysis by \citet{McQuillan2014}, which only retrieved \prot\ for 57 targets (M and K dwarfs) fainter than 16 mag. 

\subsection{\textit{Gaia} binary candidates}

In this section, we compare our target sample with the binary candidates proposed in \citet{Simonian2019} and \citet{Berger2018}, both of which used information from \textit{Gaia} DR2. Only 198 targets are in common with the \citet{Simonian2019} sample, which is focused only on the fast rotators ($P_\text{rot}<7$ d), while 22,378 targets are common to the \citet{Berger2018} sample. Appendix~\ref{app} describes these targets in \logg-\teff, \prot-\teff, \sph-\teff, and \sph-\prot diagrams.

\begin{figure}[h]\centering
\includegraphics[trim=0 0 0 29mm,clip,width=\hsize]{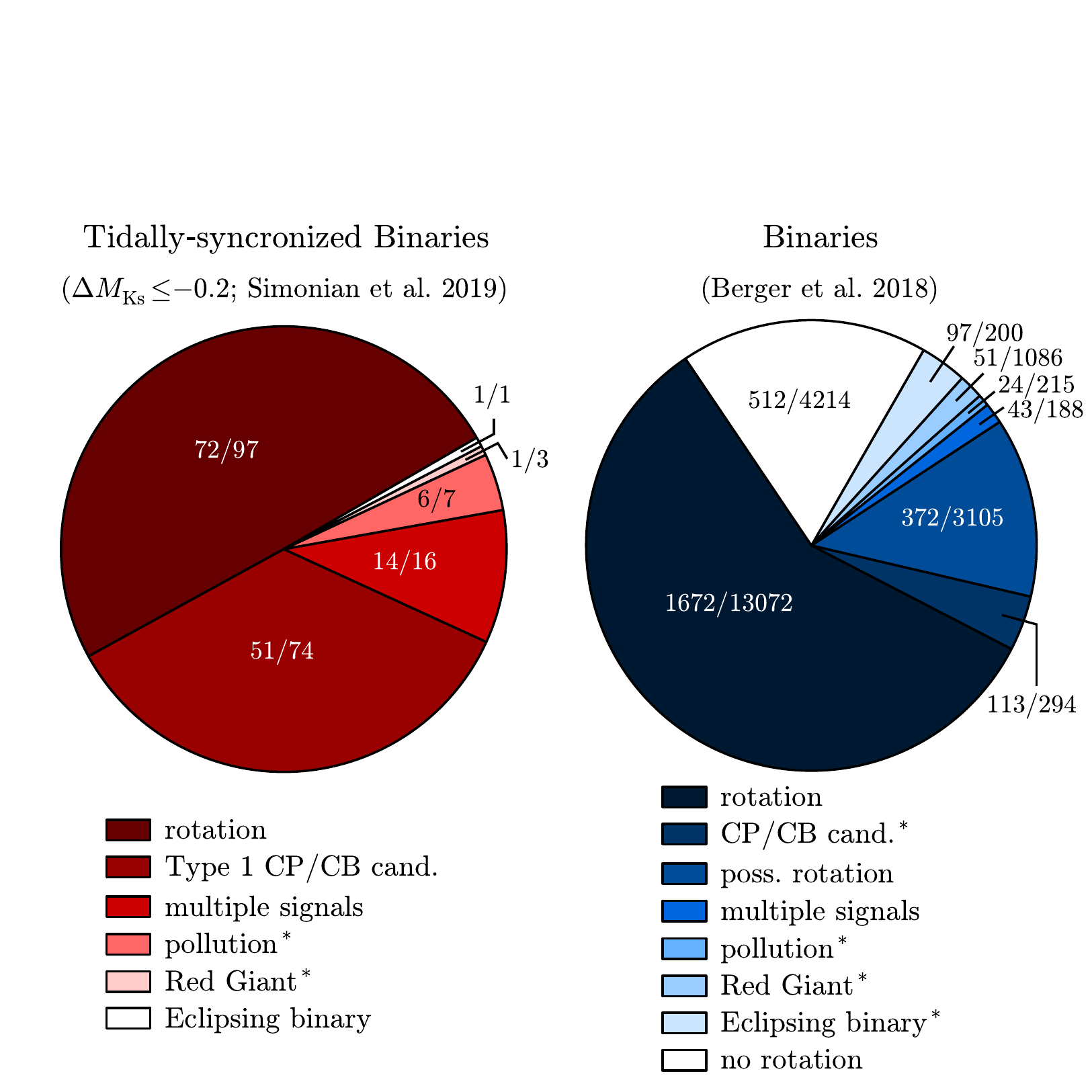}
\caption{Summary of our results for targets identified as binary candidates by \citet[][left]{Simonian2019} and \citet[][right]{Berger2018}. The size of the slices only concern the targets that are binary candidates. The annotations indicate the fraction of targets flagged as binary candidate over the total number of common targets in each category. Asterisks mark categories with sub-categories (see caption of Tables~3 and~4). Note that three RR Lyrae are analyzed and identified as single main-sequence stars by \citet{Berger2018}.}\label{fig:chartbin}
\end{figure}

Using \textit{Gaia} DR2, \citet{Simonian2019} found that faster rotators are often systematically offset in luminosity from the single-star main-sequence in comparison to slower rotators. This was interpreted as a signature of tidally-synchronized binaries, for which tidal interactions synchronize the rotation and orbital periods. Both because the fast rotator population in \citet{Simonian2019} was dominated by binary systems, and because our rapid rotators do not behave like typical active spotted stars, we advise caution in the interpretation of measurements of rapidly rotating stars.
The left-hand piechart of Fig.~\ref{fig:chartbin} summarizes the comparison between targets that are both in our sample and that of \citet{Simonian2019}. The size of the slices indicate the percentage of possible tidally-synchronized binaries of each sub-category distinguished in this work. The fractions denoted along the chart indicate the number of possible binaries over the total number of common targets between the two analyses for each sub-category. For example: seven misclassified red giants were analyzed by \citet{Simonian2019}, six of which have luminosity excess consistent with binarity, representing $\sim4\%$ of the targets in this study that are identified as synchronized binaries by \citet{Simonian2019}. $\Delta M_\text{Ks}$ indicates the luminosity excess correction (also listed in Tables~3-~5) which corresponds to the difference between the observed luminosity of a star based on the absolute magnitude in the Ks-band and the expected luminosity for a single star with a given temperature, metallicity, and age inferred from models \citep[for details see][]{Simonian2019}. We adopted the inclusive binary threshold $\Delta M_\text{Ks}<-0.2$ defined by \citet{Simonian2019}.

Interestingly, a significant fraction of possible tidally-synchronized binaries show multiple signals in their KEPSEISMIC and PDC-MAP light curves, and most of the Type~1 CP/CB candidates are identified as possible binaries in \citet{Simonian2019}. Also, four of the misclassified red giants identified as possible binaries show signatures consistent with the Type~1 CP/CB candidates. 72 targets for which we provide \prot\ estimate (none of which are Type~1 CP/CB candidates or target with multiple signals) are likely to be tidally synchronized binaries according to \citet{Simonian2019}. Note that Figs.~\ref{fig:hist}-\ref{fig:protsph} do not include binary candidates. Moreover, we do not find any particular \prot\ or \sph\ trend as a function of the luminosity excess correction. However, most of the targets that are possibly tidally-synchronized binaries ($\sim 70\%$) have \sph\ larger than $10^4$ ppm. 

Using \textit{Gaia} DR2, \citet{Berger2018} revised the Radii of the \kep\ targets and identified misclassified targets (possible subgiants and red giants) and possible binary systems. Flags are added to Tables~3-5. Most of the targets with \prot\ estimates that were not CP/CB candidates analyzed by \citet[][11,400 out of 13,072]{Berger2018} are found to be likely single stars. 113 CP/CB candidates are identified as possible binaries with 112 of those being Type~1 CP/CB candidates. 69 of the binary candidates show multiple signals in the PDC-MAP and KEPSEISMIC time-series. 101 eclipsing binaries (four are also flagged as misclassified red giants) are found to be \textit{Gaia} binary candidates. Note that misclassified red giants can be in binary systems and, thus, it is reasonable having misclassified red giants with more than one flag. The three RR Lyrae in the sample are common to the analysis of \citet{Berger2018} which identified them as single main-sequence stars.

Finally, for targets in common between \citet{Berger2018} and \citet{Simonian2019}, their results agree reasonably well. All the common targets found to be likely binary systems by \citet{Berger2018} are also identified as possible tidally-synchronized binaries by \citet{Simonian2019}. However, some of the single stars from \citet{Berger2018} are below the threshold imposed by \citet{Simonian2019} for targets to be flagged as binaries.

As mentioned in Sect.~\ref{sec:sample}, 368 presumably fast rotators do not behave as typical active stars. While we have identified three types of possible classical pulsators, it is not clear whether they are indeed classical pulsators. The results from \citet{Simonian2019} and \citet{Berger2018} suggest the interesting possibility of these Type~1 targets being close-in binary systems. A detailed analysis of these targets is beyond the scope of the current work. Nevertheless, we consider that one should be careful drawing conclusions based on the rotation estimate for fast rotators. Note that flags with the results of \citet{Simonian2019} and \citet{Berger2018} are added to Tables~3-5.

Finally, using \textit{Gaia} DR2, \citet{Berger2018} also identified the evolutionary stage of \kep\ targets. We have removed the misclassified red-giant candidates from the rotation analysis (Sect.~\ref{sec:sample}; Table~4; Garc\'ia et al. in prep). However, we did not remove the subgiant candidates from the analysis. 61 targets in Table~3 were flagged as subgiants by \citet{Berger2018}.  These targets were neglected in Figs.~\ref{fig:hist}-\ref{fig:protsph} and the \textit{Gaia} subgiant flag is provided in Tables~3 and 5.  

\section{Summary and conclusions}\label{sec:conclusion}

One can learn about surface rotation and magnetic activity by studying the brightness variations due to dark spots rotating across the visible stellar disk. In this work, we analyze \kep\ long-cadence data of 26,521 M and K main-sequence stars. The main goal of this work was to determine the average surface rotation and photometric activity level of the targets using the longest time-series available.

Rotation estimates are obtained by combining wavelet analysis, autocorrelation function, and composite spectrum of light curves \citep[e.g.][]{Mathur2010b,Garcia2014,Ceillier2016,Ceillier2017}. This methodology was found to be the best in terms of completeness and reliability \citep{Aigrain2015}. We compared the results for three KEPSEISMIC time-series (obtained with 20-day, 55-day, and 80-day filters) and PDC-MAP time-series to determine reliable rotation periods.

Given the rotation period, we also calculated the photometric activity proxy \sph\, which corresponds to the average standard deviation computed over subseries of length $5\times P_\text{rot}$ \citep{Mathur2014}. \sph\ is sensitive to the spot visibility and, thus, to their latitudinal distribution and stellar inclination angle. For this reason, \sph\ is likely to be a lower limit of the true photometric activity level. Also, in cases where spots are approximately in anti-phase (approximately $180^\circ$ apart in longitude), \sph\ will underestimate the true activity level. Nevertheless, \sph\ was demonstrated to be consistent with other solar activity proxies \citep{Salabert2017} and complementary to the chromospheric activity $S$ index for solar analogs \citep{Salabert2016a}.

We successfully recovered the surface rotation periods and respective photometric activity proxy for 15,290 stars ($\sim 62\%$ of the targets analyzed in Sect. \ref{sec:method}). We provide period estimates for targets whose KEPSEISMIC and PDC-MAP light curves show multiple signals (270 targets). We also provide period estimates for another 350 stars that we flagged as possible classical pulsators or close-in binary systems. Their behaviour is not consistent with that of single active stars, resembling that of RR Lyrae or Cepheids. We also have identified $\gamma$-Doradus or $\delta$-Scuti candidates (18 in total). We note, however, that further analysis is needed to properly classify these 368 targets and determine the source of the multiple signals in the light curves of the 270 targets.

Another 3,562 targets ($\sim14\%$ of the sample) show spot modulation in their light curves, but we are unable recover reliable rotation periods. 5,310 targets ($\sim20\%$ of the sample) do not exhibit any apparent spot modulation. The magnitude distribution of these targets is slightly shifted towards fainter values in comparison with stars with spot modulation in the light curves. We do not provide rotation estimates for confirmed RR Lyrae (3 stars; Szab\'o et al. in prep), known eclipsing binaries \citep[272 stars;][]{Kirk2016,Adbul-Masih2016}, targets identified as misclassified red giants (1221; Garc\'ia et al. in prep), and targets whose light curves show evidence for photometric pollution (255 targets). We consider a light curve photometrically polluted when only particular \kep\ Quarters show modulation signals or the signal is only present in the KEPSEISMIC light curves. These targets are listed in Table~4.

\citet{Berger2018} and \citet{Simonian2019} identified possible binary systems and we have crossed-checked our sample with their results. In terms of rotation and photometric activity proxy, we did not find any particular difference between binaries and single stars. Nevertheless, it is interesting to note that a significant number of targets show evidence of photometric pollution by nearby stars. Also, most of the classical pulsator candidates flagged as binary candidates show stable, high-amplitude variations and beating patterns, and we therefore treat them as classical pulsator/close-in binary candidates. Note that we did not remove binary candidates from the analysis, but did include the respective flags from \citet{Berger2018} and \citet{Simonian2019} in our tables.

Only $\sim72\%$ of the targets with rotation period estimates are also detected in spot modulation in \citet{McQuillan2013a,McQuillan2014}. For the common targets, the agreement on the \prot\ estimate is about $99.4\%$ at $2\sigma$. We also show that our methodology is able to recover rotation periods for a larger number of stars (4,431 additional \prot) than the analysis by \citet{McQuillan2014}. In particular, we provide \prot\ for a higher fraction of cool and faint stars.

For the parameter range studied here (M and K~dwarfs), we find that the mean rotation period to decrease with increasing stellar effective temperature and mass, i.e. K~dwarfs are on average faster rotators than M~dwarfs. This is consistent with previous findings \citep[e.g][]{McQuillan2014,Garcia2014}. As in \citet{McQuillan2014}, we also found two sequences in the \prot-\teff\ relation: a wider and more populated sequence for slower rotators and a narrower and less populated sequence for faster rotators. The bimodality is clear in the rotation-period distribution for M~dwarfs. Due to the wider range of effective temperatures of K~dwarfs in comparison with the M~dwarfs in the sample, the bimodality is not clear in the \prot\ distribution for K~dwarfs. However, we verified that the bimodality is present while splitting the K~dwarfs in smaller sub-samples according to their temperature. Furthermore, the bimodality is also visible in the density plot of rotation period as a function of effective temperature. 

For M and K~dwarfs, we found that the photometric activity proxy takes on a wider range of values as effective temperature and mass increase, and the extremes of the distribution extend to both higher and lower \sph\ values.

The photometric activity proxy \sph\ increases as rotation period decreases. This is consistent with faster rotators being more active than slower rotators \citep[e.g.][]{Vaughan1981,Baliunas1983,Noyes1984b}. The bimodal rotation-period distribution is also visible through the two branches in the \sph-\prot\ relation. A similar behaviour was also found by \citet{McQuillan2013,McQuillan2014} while using a different measure of photometric variability, $R_\text{var}$ \citep[see Sect. \ref{sec:sph};][]{Basri2011,Basri2013}.

Based on the evidence of two distinct proper motion distributions, \citet{McQuillan2013} interpreted the bimodal rotation-period distribution as evidence for two stellar populations with different ages associated to different star-formation episodes. Using \textit{Gaia} data, the results by \citet{Davenport2017} and \citet{Davenport2018} are consistent with the bimodal rotation-period distribution being associated to a bimodal age distribution. In particular, the authors found that the bimodality is more pronounced at low Galactic scale height which is assumed to be an age indicator. \citet{Montet2017} and \citet{Reinhold2019} found that the fast rotating, more active sequence corresponds to spot-dominated stars, while the slowly rotating less active stars are faculae-dominated. These studies support the idea that solar-type stars transition from spot-dominated to faculae-dominated as stars evolve. \citet{Reinhold2019} suggested that the observed period bimodality is actually a dearth of detections at intermediate rotation periods due to the cancellation between dark spots and bright faculae.
In this work, we found that the photometric activity proxy \sph\ varies approximately within the same range for M~dwarfs in both fast and slow rotator branches. For K~dwarfs, although most of the targets in both branches have \sph\ values smaller than $\sim 7000$~ppm, \sph\ values for slow rotators extend to significantly smaller values ($\sim 200$~ppm), while the fast rotators are mostly within $\sim 600-7000$~ppm. 

The methodology followed in this work will be extended to G and F main-sequence stars and subgiants in a future paper. See \citet{Santos2018b} for a brief summary where the analysis is also applied to G main-sequence stars cooler than 5500 K and subgiants cooler than 5500 K and with surface gravities larger than $\log\,g=3.5$ dex.

\acknowledgments
The authors thank R\'obert Szab\'o, Paul G. Beck, Katrien Kolenberg, and Isabel L. Colman for helping on the classification of stars. This paper includes data collected by the \kep\, mission and obtained from the MAST data archive at the Space Telescope Science Institute (STScI). Funding for the \kep\ mission is provided by the National Aeronautics and Space Administration (NASA) Science Mission Directorate. STScI is operated by the Association of Universities for Research in Astronomy, Inc., under NASA contract NAS 5–26555. ARGS acknowledges the support from NASA under Grant NNX17AF27G. RAG and LB acknowledge the support from PLATO and GOLF CNES grants. SM acknowledges the support from the Ramon y Cajal fellowship number RYC-2015-17697. TSM acknowledges support from a Visiting Fellowship at the Max Planck Institute for Solar System Research. This research has made use of the NASA Exoplanet Archive, which is operated by the California Institute of Technology, under contract with the National Aeronautics and Space Administration under the Exoplanet Exploration Program.

\software{KADACS \citep{Garcia2011}, NumPy \citep{numpy}, SciPy \citep{scipy}, Matplotlib \citep{matplotlib}}

\facility{MAST, \kep\ Eclipsing Binary Catalog, Exoplanet Archive}

\bibliographystyle{aasjournal}
\bibliography{rotationbib}\pagebreak\clearpage

\begin{figure*}[h]
\includegraphics[trim=0 20 0 0mm,clip,angle=90,width=\hsize]{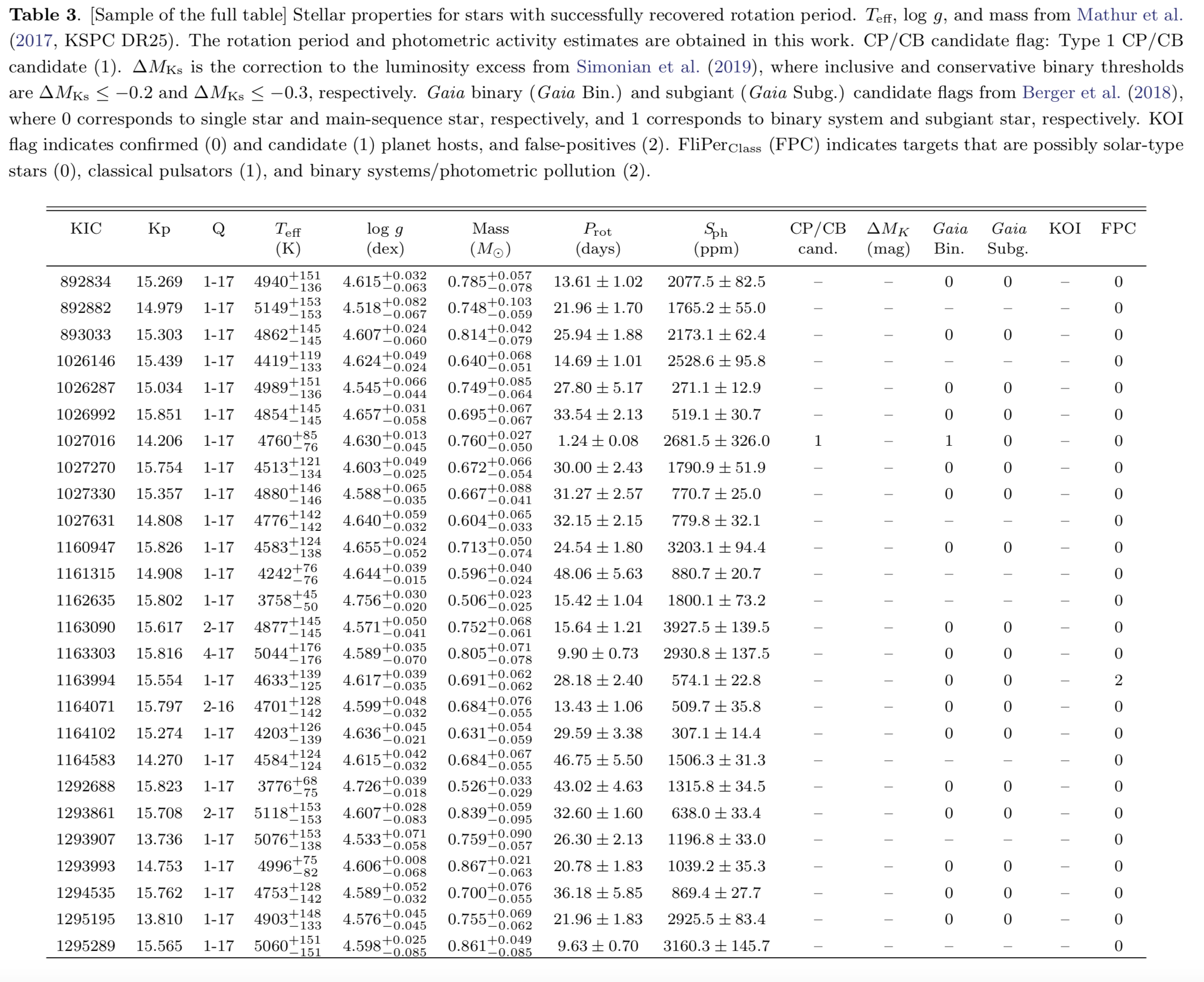}
\end{figure*}\pagebreak\clearpage

\begin{figure*}[h]\centering\vspace{3cm}
\includegraphics[trim=0 20 0 0mm,clip,angle=90,width=\hsize]{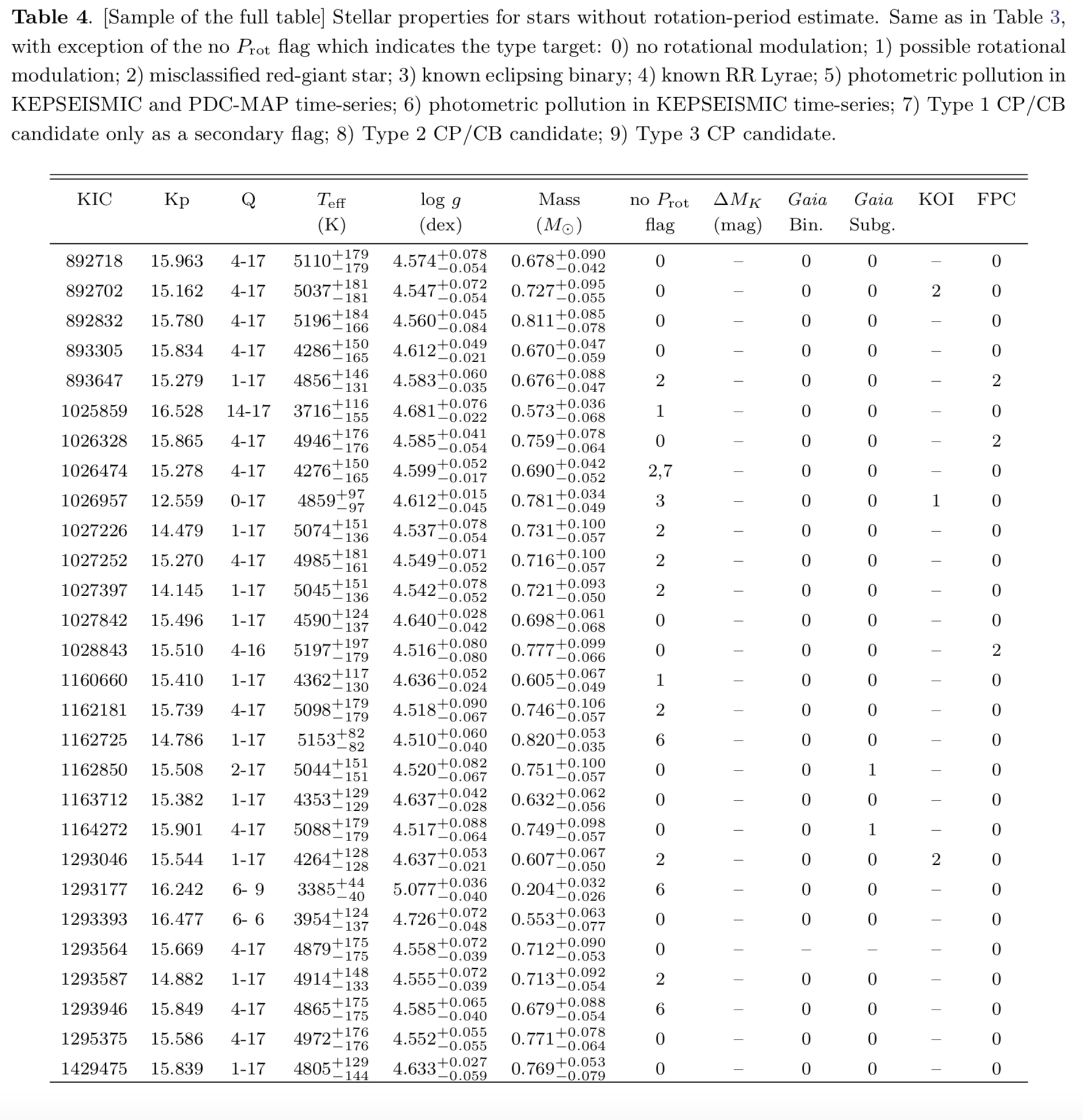}
\end{figure*}\pagebreak\clearpage

\begin{figure*}[h]\hspace{1.2cm}\vspace{-1.3cm}
\includegraphics[trim=0 20 0 0mm,clip,angle=90,width=0.9\hsize]{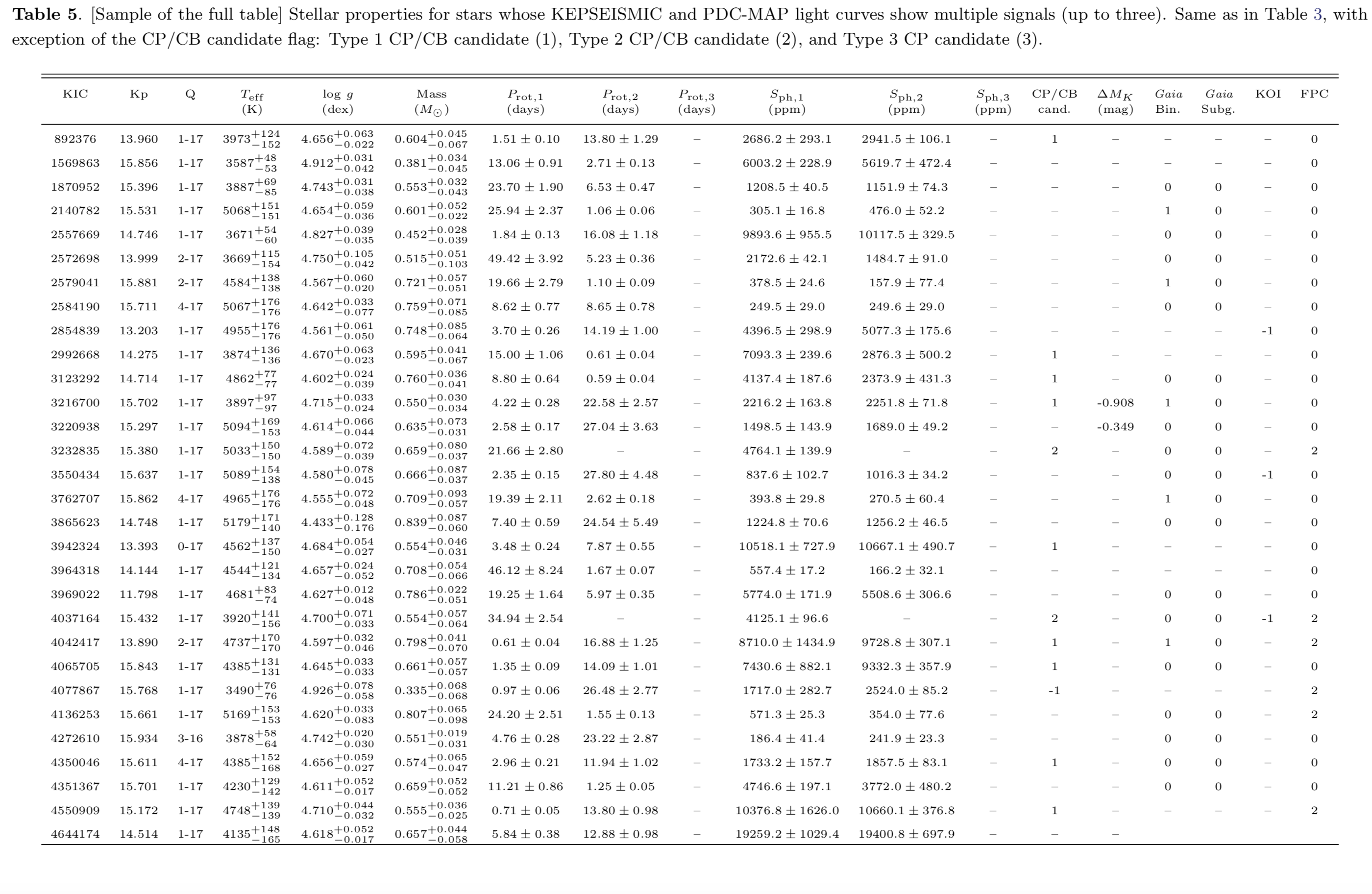}
\end{figure*}\pagebreak\clearpage

\appendix
\section{Rotation and photometric activity index including potential non-single non-main-sequence M and K stars}\label{app}

The study presented here is focused on main-sequence M and K stars selected according to the \kep\ Stellar Properties Catalog for Data Release 25 \citep[KSPC DR25;][]{Mathur2017}. However, there is a number of possible polluters in the sample, as well as potential binary systems.

Here, we do not perform the rotation analysis for misclassified red giants (Garc\'ia et al. in prep.), confirmed RR Lyare (Szab\'o et al. in prep.), eclipsing binaries \citep[Villanova \kep\ Eclipsing Binary Catalog;][see \citet{Lurie2017} for rotational analysis of these systems]{Kirk2016,Adbul-Masih2016}, and and Type~2 and 3 CP/CB candidates. In addition to these polluters, we have identified other potential non-single non-main-sequence M and K stars, including those flagged by \citet{Berger2018}, \citet{Simonian2019}, and FliPer$_\text{Class}$ \citep{Bugnet2019}. In this section, we add the results from the rotational analysis and photometric activity index of these targets. Note that the results are listed in the Tables~3-5 with the respective flags.

\begin{figure*}[htp]\centering
\includegraphics[width=\hsize]{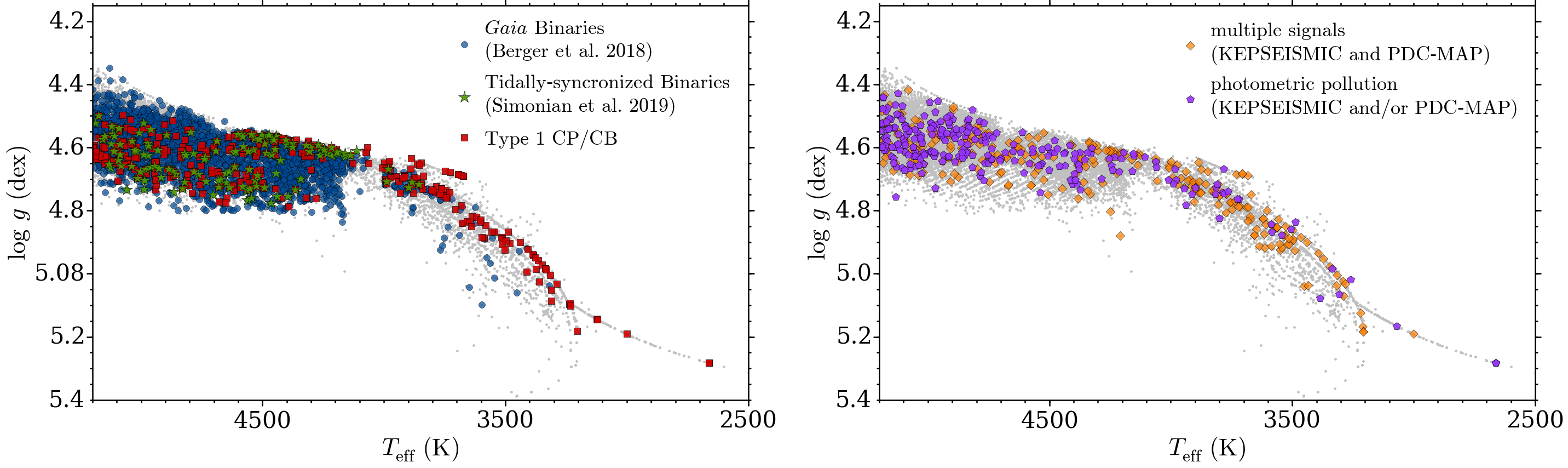}
\caption{Surface gravity-effective temperature diagram for potential non-single non-main-sequence M and K stars. Effective temperature (\teff) and surface gravity (\logg) values are adopted from KSPC DR25. For context, the remainder of the sample is shown in gray. Blue and red symbols show binary systems flagged by \citet{Berger2018} and \citet{Simonian2019}, respectively. Green, orange, and purple mark the targets flagged as Type~1 CP/CB candidates, showing multiple signals, and photometric pollution, respectively.} \label{fig:hrbin}
\end{figure*}

Figure~\ref{fig:hrbin} shows the surface gravity-effective temperature diagram for some of the potential non-single M and K stars. Targets in our sample that were flagged as binaries by \citet[][2,841 targets]{Berger2018} are shown in blue, while tidally-synchronized binaries identified by \citet{Simonian2019} are shown in red. The green, orange, and purple symbols indicate targets that were flagged in this study as Type~1 CP/CB candidates, showing multiple signals (in both KEPSEISMIC and PDC-MAP light curves), or being 
photometrically polluted. Multiple signals and photometric pollution may result from nearby stars in the field of view or from different components of multiple systems. The apertures used for KEPSEISMIC light curves are larger than those used for PDC-MAP. Therefore, a given signal that only exists in KEPSEISMIC light curves is likely due to photometric pollution by a nearby star in the field of view, while multiple signals present in both KEPSEISMIC and PDC-MAP light curves are likely associated to unresolved sources. Determining whether the multiple signals are due to background stars or actual binary/multiple systems is beyond the scope of this study. Nevertheless, we determine the periods of the signals (Table~5).

\begin{figure}[h]\centering
\includegraphics[width=0.5\hsize]{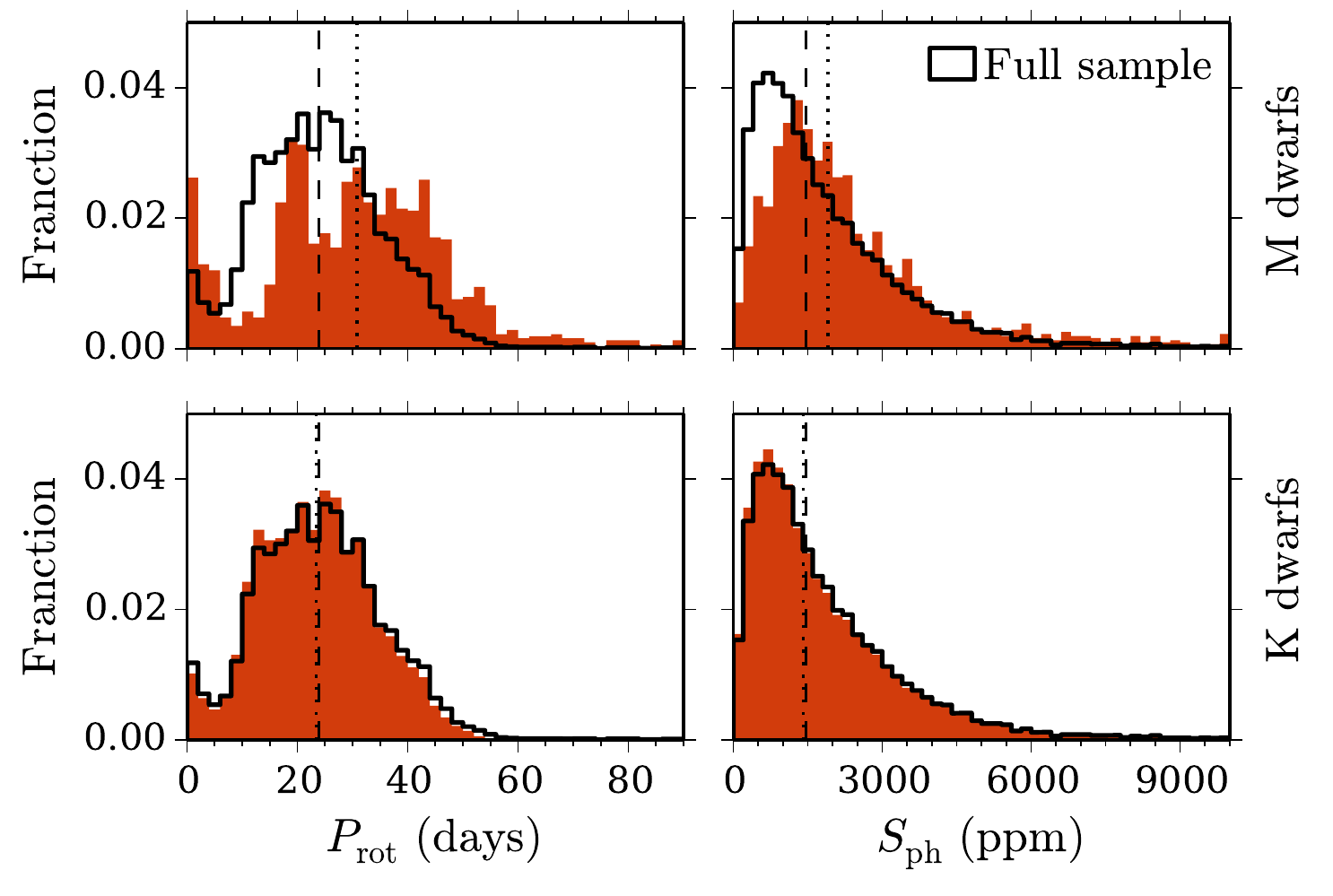}
\caption{Same as in Fig.~\ref{fig:hist} but also including potential non-single non-main-sequence stars.}\label{fig:histapp}\vspace{0.3cm}
\end{figure}

\begin{figure*}[htp]
\includegraphics[width=\hsize]{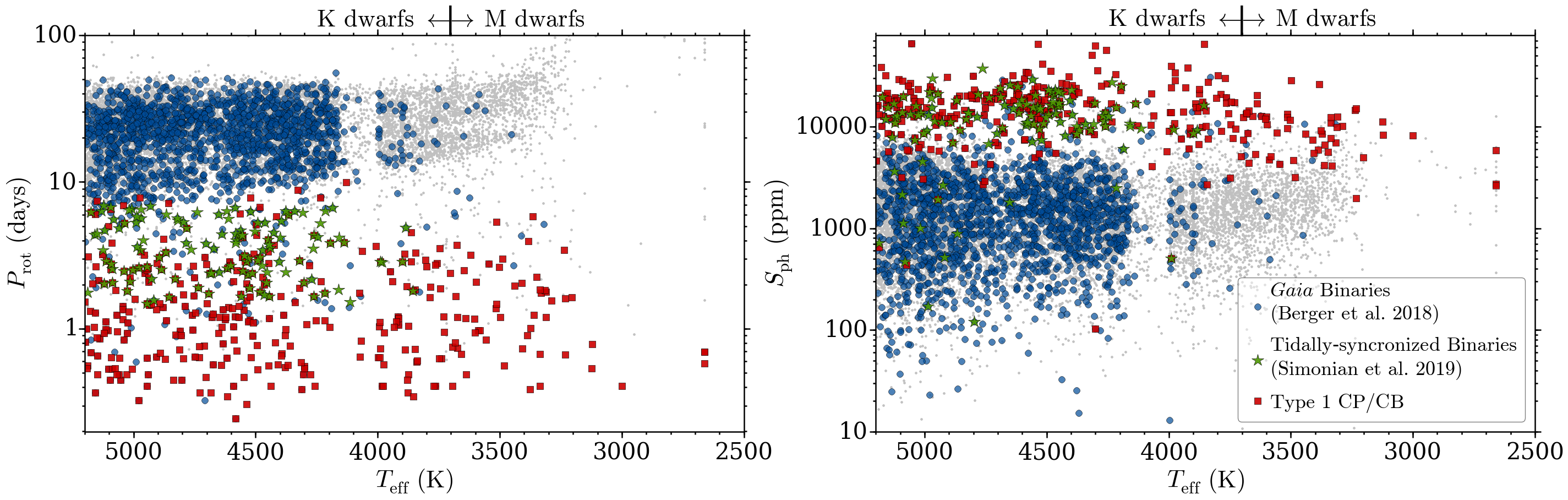}
\caption{Same as left-hand panels of Figs.~\ref{fig:protTeffMass} and \ref{fig:sphTeffMass} but including binaries identified by \citet[][blue circles]{Berger2018}, tidally-synchronized binaries identified by \citet[][green stars]{Simonian2019}, and Type~1 CP/CB candidates identified in this work (red squares). For reference the single M and K dwarfs are marked in gray.} \label{fig:protsphTeffbin}\vspace{0.3cm}
\end{figure*}

\begin{figure*}[htp]\vspace{-0.2cm}
\includegraphics[width=\hsize]{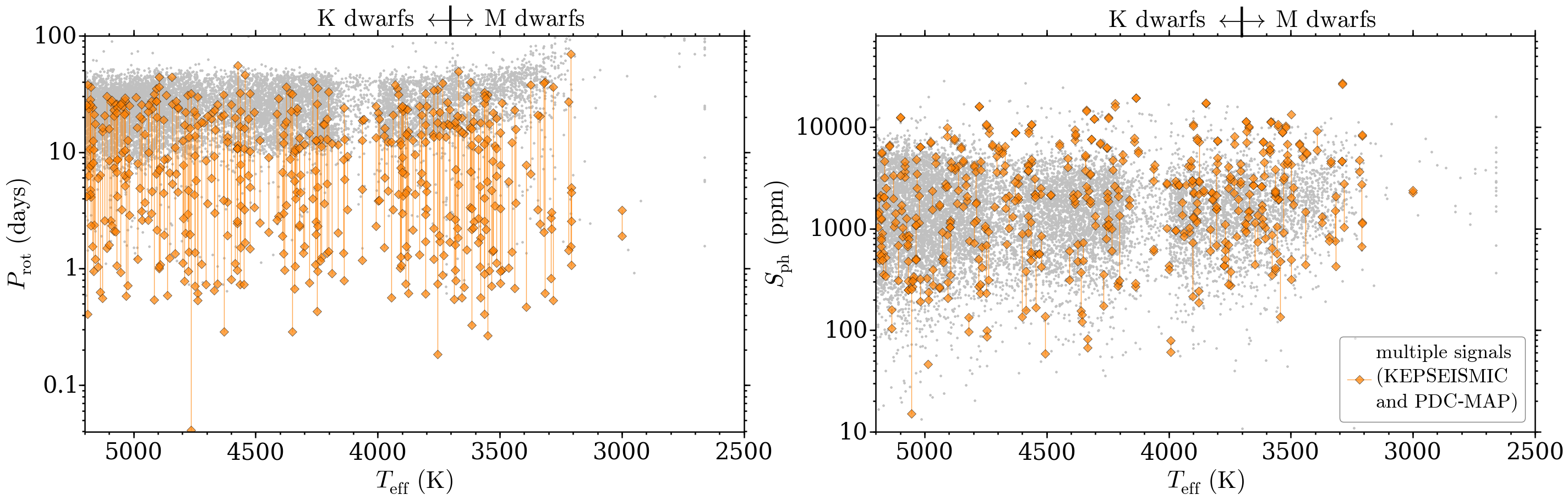}\vspace{-0.2cm}
\caption{Same as left-hand panels of Figs.~\ref{fig:protTeffMass} and \ref{fig:sphTeffMass} but including the targets whose light curves show multiple signals (orange diamonds). For reference the single M and K dwarfs are marked in gray. For illustration purpose the \prot\ axis is different in this figure.} \label{fig:protsphTeffmsig}\vspace{-1cm}
\end{figure*}

\begin{figure}[h]\centering
\includegraphics[width=0.5\hsize]{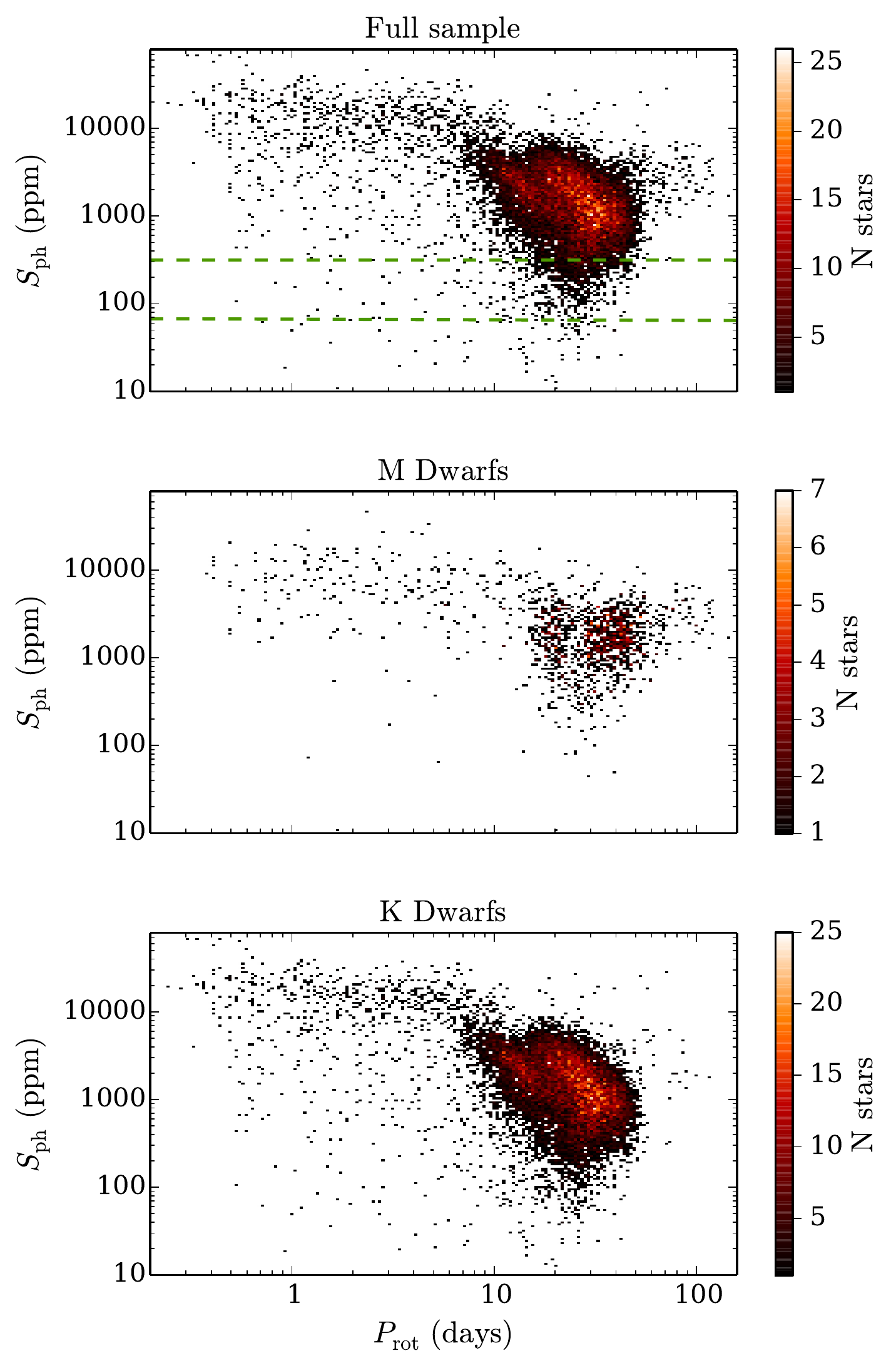}\vspace{-0.3cm}
\caption{Same as in Fig.~\ref{fig:protsph} but also including potential non-single non-main-sequence stars.}\label{fig:protsphapp}\vspace{-2cm}
\end{figure}

\begin{figure*}[htp]\centering
\includegraphics[width=0.7\hsize]{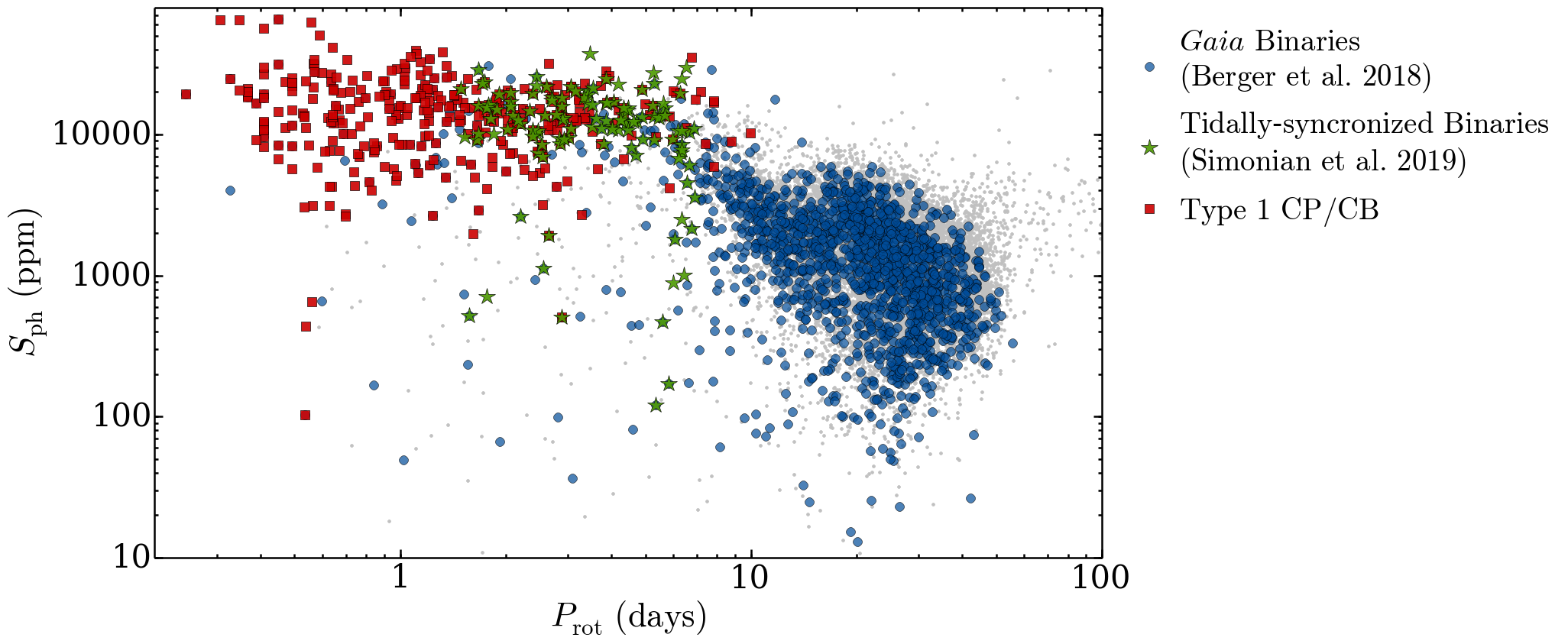}
\caption{Same as in Fig.~\ref{fig:protsph} but including binaries identified by \citet[][blue circles]{Berger2018}, tidally-synchronized binaries identified by \citet[][green stars]{Simonian2019}, and Type~1 CP/CB candidates identified in this work (red squares). For reference the single M and K dwarfs are marked in gray.} \label{fig:protsphbin}
\end{figure*}

Figure~\ref{fig:histapp} shows the \prot\ and \sph\ distributions when including the potential non-single non-main-sequence stars. Compared with Fig.~\ref{fig:hist}, which only represents stars that are likely single main-sequence stars, there is a clear increase of targets at the fast-rotator regime. In particular, there is a significant number of targets with a large photometric activity proxy \sph . This is also seen in Figs.~\ref{fig:protsphTeffbin}-\ref{fig:protsphbin}.

\begin{figure*}[htp]
\includegraphics[width=0.5\hsize]{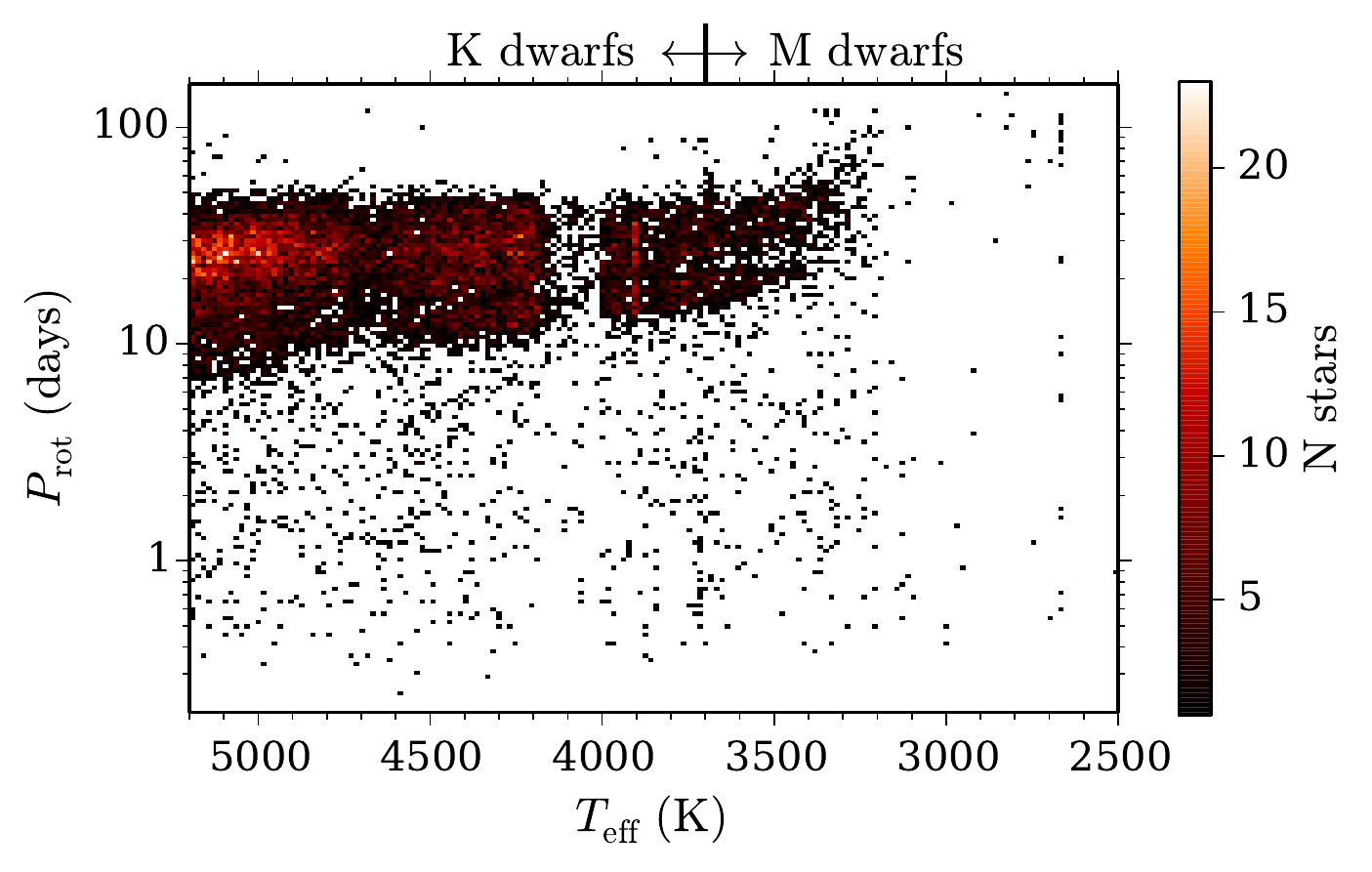}
\includegraphics[width=0.524\hsize]{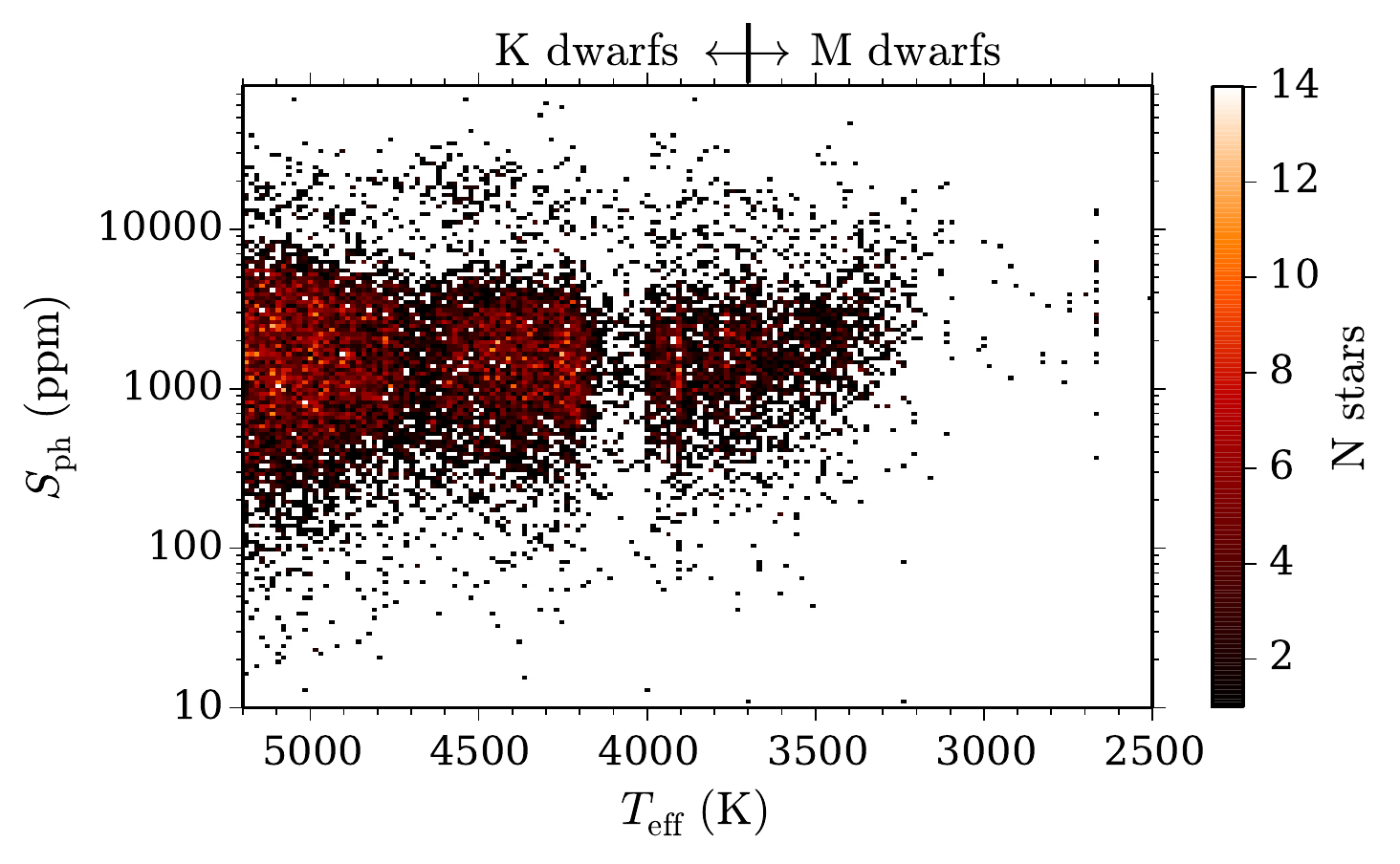}
\caption{Same as left-hand panels of Figs.~\ref{fig:protTeffMass} and \ref{fig:sphTeffMass} but also including potential non-single non-main-sequence stars.} \label{fig:protsphTeffapp}\vspace{0.3cm}
\end{figure*}

Figures \ref{fig:protsphTeffbin}, \ref{fig:protsphTeffmsig}, and \ref{fig:protsphbin} show the \prot-\teff, \sph-\teff, and \sph-\prot\ diagrams where we mark the individual groups of potential binaries, Type~1 CP/CB candidates, and targets whose KEPSEISMIC and PDC-MAP light curves show multiple signals. Most of the binaries identified by \citet{Berger2018} with \prot\ estimate (blue circles; 1,784 targets in Table~3) occupy in general the same parameter space as single M and K dwarfs (gray; see also Figs.~\ref{fig:protTeffMass}-\ref{fig:protsph}). \citet{Simonian2019} focused on fast rotators identifying targets that are likely tidally-synchronized binary systems (green symbols; 123 targets in Table~3). These targets have shorter periods and tend to have larger \sph\ values than most of the single M and K dwarfs. Type~1 CP/CB candidates (red squares; 350 targets in Table~3), identified in this work, occupy approximately the same parameter space as the tidally-synchronized binaries. Figure~\ref{fig:protsphTeffmsig} shows the results for the 270 light curves with multiple signals (Table~5). For most of the targets the period of at least one of the signals is below the \prot-\teff\ relation followed by most of the single M and K stars. The  \sph values reflect the contribution from multiple signals and therefore should be used with care.

\section{Comparison with the rotation results for PDC-MAP and KEPSEISMIC data sets}\label{app2}

\begin{figure}[h]\centering
\includegraphics[width=0.5\hsize]{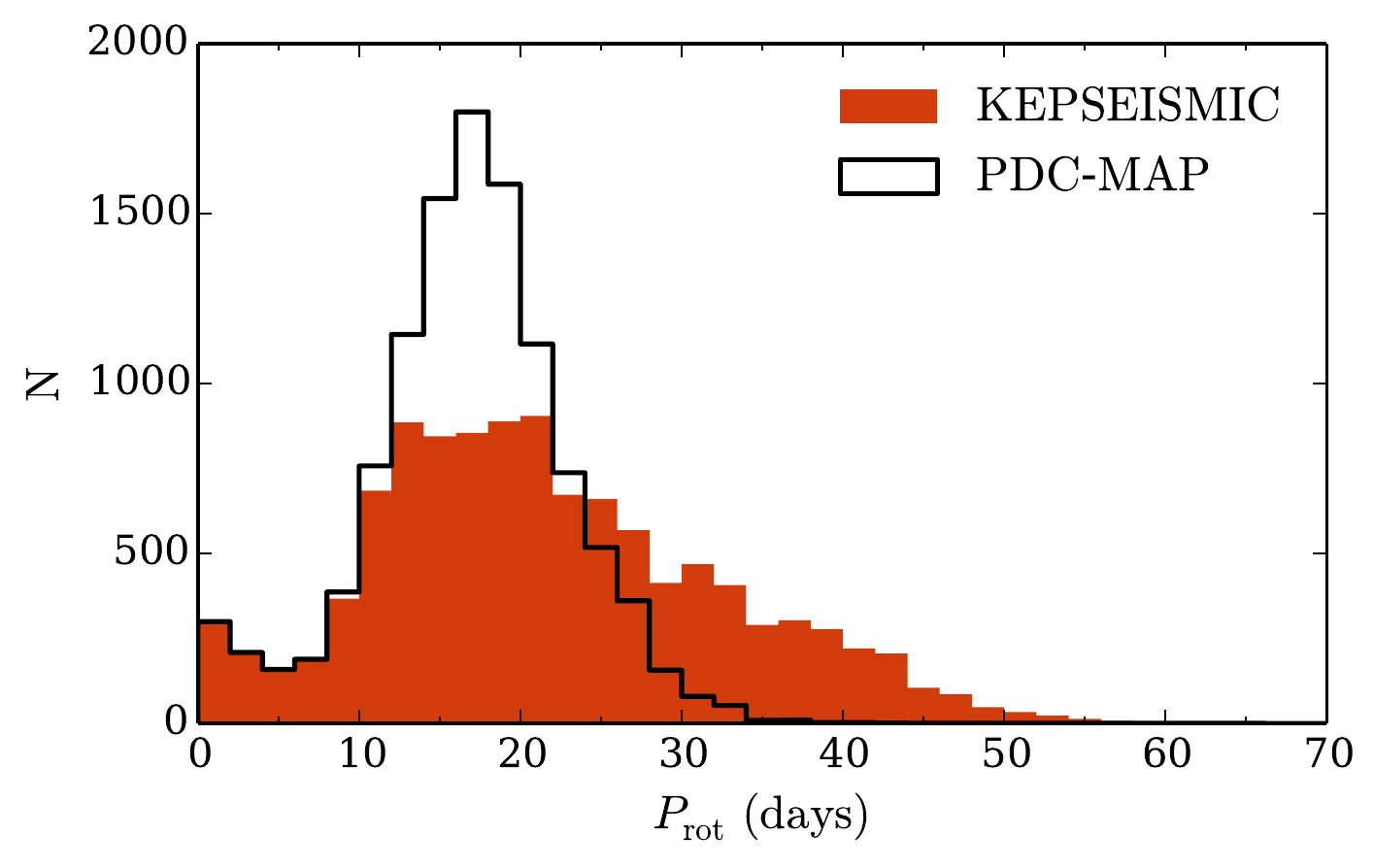}
\caption{Rotation-period distributions for KEPSEISMIC (red) and PDC-MAP (black) light curves. Only the common automatically selected targets (11,131 targets) are represented.}\label{fig:pdc}
\end{figure}

In this section, we present the comparison between the rotation periods obtained from KEPSEISMIC light curves with those obtained from PDC-MAP light curves for DR 25 (Fig.~\ref{fig:pdc}). Only the common automatically selected targets are represented. Since we only have one PDC-MAP data set, the step in Sect.~\ref{sec:auto} concerning the multiple filters is skipped for PDC-MAP. This comparison illustrates how biased the rotation estimates would be by adopting PDC-MAP light curves, which leads to a prominent peak around 17 days with very few rotation periods longer than 30 days. Only for $72.7\%$ of targets, the rotation-period estimates from KEPSEISMIC and PDC-MAP agree. With the PDC-MAP light curves, we retrieve the second harmonic ($1/2P_\text{rot}$) for $25.1\%$ of the targets and the third harmonic ($1/3P_\text{rot}$) for $1.3\%$ of the targets. PDC-MAP calibration and filtering are not uniform for all \kep\ Quarters or all targets. Also, a high-pass filter of 20 days is often applied for PDC-MAP. We have produced our own light curves (KEPSEISMIC) with customized apertures and a uniform calibration through the implementation of KADACS \citep{Garcia2011}. Gaps shorter than 20 days are filled by applying in-painting techniques \citep{Garcia2014a,Pires2015}. Furthermore, we obtain three different KEPSEISMIC data sets with high-pass filters of different cut-off periods. This allows us to identify the true rotation period instead of one of its harmonics. Therefore, besides being optimized for seismic studies, KEPSEISMIC light curves are also more adequate for rotation studies.

\end{document}